\documentclass[paper]{JHEP3}
\usepackage{epsfig}
\usepackage{amsmath}
\usepackage{amssymb}
\usepackage{amsbsy}
\usepackage{enumerate}
\usepackage{bbm}
\usepackage{url}

\usepackage[hang,sf]{subfigure}

\newcommand{\tmmathbf}[1]{\ensuremath{{\bf #1}}}

\def\beq{\begin{equation}}
\def\beqn{\begin{eqnarray}}
\def\eeq{\end{equation}}
\def\eeqn{\end{eqnarray}}
\def\abs#1{\left|#1\right|}

\newcommand\FKS{Frixione, Kunszt and Signer}

\newcommand\HERWIG{{\tt HERWIG}}

\newcommand\PYTHIA{{\tt PYTHIA}}

\newcommand\MINT{{\tt MINT}}

\def\lq{\left[} 
\def\rq{\right]} 
\def\rg{\right\}} 
\def\lg{\left\{} 
\def\({\left(} 
\def\){\right)}

\newcommand\sss{\mathchoice%
{\displaystyle}%
{\scriptstyle}%
{\scriptscriptstyle}%
{\scriptscriptstyle}%
}

\newcommand\nplus{\oplus}
\newcommand\nminus{\ominus}

\newcommand\splus{{\sss \nplus}}
\newcommand\sminus{{\sss \nminus}}

\newcommand\splusminus{{\mathchoice%
{\vplusminus\displaystyle}%
{\vplusminus\scriptstyle}%
{\vplusminus\scriptscriptstyle}%
{\vplusminus\scriptscriptstyle}%
}}

\newcommand\sminusplus{{\mathchoice%
{\vminusplus\displaystyle}%
{\vminusplus\scriptstyle}%
{\vminusplus\scriptscriptstyle}%
{\vminusplus\scriptscriptstyle}%
}}

\newdimen\hbigcirc
\newdimen\wbigcirc

\newdimen\figwidth

\newcommand\captskip{\vskip -0.7cm}

\figwidth=\textwidth

\newcommand\vplusminus[1]{%
\settoheight{\hbigcirc}{$#1\bigcirc$}%
\settowidth{\wbigcirc}{$#1\bigcirc$}%
\makebox[\wbigcirc]{%
\makebox[0pt]{\rule[0.4\hbigcirc]{0.5\wbigcirc}{0.05\hbigcirc}}%
\makebox[0pt]{\rule[0.1\hbigcirc]{0.5\wbigcirc}{0.05\hbigcirc}}%
\makebox[0pt]{\rule[0.1\hbigcirc]{0.05\wbigcirc}{0.6\hbigcirc}}%
\makebox[0pt]{$#1\bigcirc$}}%
}

\newcommand\vminusplus[1]{%
\settoheight{\hbigcirc}{$#1\bigcirc$}%
\settowidth{\wbigcirc}{$#1\bigcirc$}%
\makebox[\wbigcirc]{%
\makebox[0pt]{\rule[0.2\hbigcirc]{0.5\wbigcirc}{0.05\hbigcirc}}%
\makebox[0pt]{\rule[0.5\hbigcirc]{0.5\wbigcirc}{0.05\hbigcirc}}%
\makebox[0pt]{\rule[-0.1\hbigcirc]{0.05\wbigcirc}{0.6\hbigcirc}}%
\makebox[0pt]{$#1\bigcirc$}}%
}

\newcommand\xplus{x_\splus}
\newcommand\xminus{x_\sminus}
\newcommand\xplusminus{x_\splusminus}

\newcommand\Kplus{K_\splus}
\newcommand\Kminus{K_\sminus}
\newcommand\bxplus{\bar{x}_\splus}
\newcommand\bxminus{\bar{x}_\sminus}
\newcommand\bxplusminus{\bar{x}_\splusminus}

\newcommand\clH{{\mathbb H}}
\newcommand\clS{{\mathbb S}}

\newcommand\as{\alpha_{\sss\rm S}}

\newcommand\Lum{{\cal L}}

\newcommand\pt{p_{\sss\rm T}}
\newcommand\ptmin{{\pt^{\min}}}
\newcommand\kt{k_{\sss\rm T}}

\newcommand\ximax{{\xi_{\sss\rm{M}}}}
\newcommand\boost{\mathbb B}

\newcommand\matR{{\cal R}}

\newcommand\ctindr{{\alpha_{\sss\rm r}}}

\newcommand\Mrec{M_{\rm rec}}

\newcommand\mmod[1]{\underline{#1}}
\newcommand\MCatNLO{{\tt MC@NLO}}

\newcommand\CF{C_{\sss\rm F}}
\newcommand\TF{T_{\sss\rm F}}

\newcommand\fb{{f_b}}

\newcommand\POWHEG{{\tt POWHEG}}

\newcommand\Rad{\Phi_{\rm rad}}

\newcommand\Radmc{\Rad^{\rm \sss MC}}

\newcommand\ISRRad{\Phi_{\rm rad}^{\sss\rm ISR}}
\newcommand\FSRRad{\Phi_{\rm rad}^{\sss\rm FSR}}
\newcommand\ktISR{k_{{\sss\rm T},{\sss \rm ISR}}}
\newcommand\ktFSR{k_{{\sss\rm T},{\sss \rm FSR}}}

\newcommand\powPS{\Phi_{\sss \rm POW}}
\newcommand\decPS{\Phi_{\sss t \to b\bar{\ell}\nu }}

\newcommand\muF{\mu_{\sss\rm F}}

\newcommand\frindsing{{\ctindr}}

\newcommand\qb{\bar{q}}

\newcount\minutes 
\newcount\scratch 
 
\def\timestamp{%
\scratch=\time 
\divide\scratch by 60 
\edef\hours{\the\scratch} 
\multiply\scratch by 60 
\minutes=\time 
\advance\minutes by -\scratch 
---$\,$\hours:\null 
\ifnum\minutes< 10 0\fi 
\the\minutes}

\title{NLO single-top production matched with shower \\ in {\tt\bf POWHEG}:
$\boldsymbol{s}$- and $\boldsymbol{t}$-channel contributions}
\author{Simone Alioli\\
  Universit\`a di Milano-Bicocca and INFN, Sezione di Milano-Bicocca\\
  Piazza della Scienza 3, 20126 Milan, Italy\\
  E-mail: \email{Simone.Alioli@mib.infn.it}}
\author{Paolo Nason\\
  INFN, Sezione di Milano-Bicocca,
  Piazza della Scienza 3, 20126 Milan, Italy\\
  E-mail: \email{Paolo.Nason@mib.infn.it}}
\author{Carlo Oleari\\
  Universit\`a di Milano-Bicocca and INFN, Sezione di Milano-Bicocca\\
  Piazza della Scienza 3, 20126 Milan, Italy\\
  E-mail: \email{Carlo.Oleari@mib.infn.it}}
\author{Emanuele Re\\
  Universit\`a di Milano-Bicocca and INFN, Sezione di Milano-Bicocca\\
  Piazza della Scienza 3, 20126 Milan, Italy\\
  E-mail: \email{Emanuele.Re@mib.infn.it}}
\abstract{
  We present a next-to-leading order calculation of single-top production
  interfaced to Shower Monte Carlo programs,
  implemented according to
  the \POWHEG{} method. A detailed comparison
  with \MCatNLO{} and \PYTHIA{} is carried out for several observables,
  for the Tevatron and LHC colliders.
}

\begin{document}

\section{Introduction}
\label{sec:introduction}
Top-quark production in hadronic collisions has been one of the most
studied signal in the last twenty years. Up to recent times,
$t\bar{t}$ pair production has been the only observed top-quark
source at the Tevatron collider, due to its large, QCD-dominated,
cross section. Processes where only one top quark appears in the final
state are known in literature as single-top processes. Their cross
sections are smaller than the $t\bar{t}$ pair one, due to their weak
nature.  This fact, together with the presence of large $W+\mbox{jet}$
and $t\bar{t}$ backgrounds, makes the single-top observation very
challenging, so that this signal has been observed only recently by
the CDF~\cite{Aaltonen:2009jj} and D0~\cite{Abazov:2009ii}
collaborations.

In spite of its relative small cross section, single-top production is an
important signal for several reasons (see also
refs.~\cite{Beneke:2000hk,Harris:2002md} and references therein). Within the
Standard Model, the single-top signal allows a direct measurement of the
Cabibbo-Kobayashi-Maskawa~(CKM) matrix element $V_{tb}$~\cite{Alwall:2006bx}
and of the $b$ parton density. Furthermore, the V-A structure of weak
interactions can be directly probed in these processes, since the top quark
decays before hadronizing, and its polarization can be directly observed in
the angular correlations of its decay
products~\cite{Mahlon:1996pn,Mahlon:1999gz}.  Finally, single-top processes
are expected to be sensitive to several kinds of new physics effects and, in
some cases, are the best channels to observe
them~\cite{Tait:2000sh,Cao:2007ea,Plehn:2009it}. For all the above reasons,
single-top is an important Standard Model processes to be studied at the LHC,
where the statistics limitation due to the small cross section is less severe
and differential distributions can also be studied.

In order to include a reliable description of both short- and long-distance
effects into the simulation of hadronic processes, it is important to
consistently match fixed order results with parton showers.  Radiative
corrections for single-top production have been known for
years~\cite{Harris:2002md,Bordes:1994ki,Stelzer:1997ns,Sullivan:2004ie,Campbell:2004ch,Campbell:2005bb,
Cao:2004ap,Cao:2005pq,Campbell:2009ss}, while the implementation of these
results into a next-to-leading-order Shower Monte Carlo~(SMC), namely
\MCatNLO{}~\cite{Frixione:2002ik,Frixione:2003ei}, is more
recent~\cite{Frixione:2005vw,Frixione:2008yi}.

In this work we present a next-to-leading order (NLO) calculation of
$s$- and $t$-channel single-top production, interfaced to Shower Monte
Carlo programs, according to the \POWHEG{} method.  This method
was first suggested in ref.~\cite{Nason:2004rx}, and was described
in great detail in ref.~\cite{Frixione:2007vw}.
Until now, the \POWHEG{}
method has been applied to $ZZ$ pair
hadroproduction~\cite{Nason:2006hfa}, heavy-flavour
production~\cite{Frixione:2007nw}, $e^+ e^-$ annihilation into
hadrons~\cite{LatundeDada:2006gx} and into top
pairs~\cite{LatundeDada:2008bv}, Drell-Yan vector boson
production~\cite{Alioli:2008gx,Hamilton:2008pd}, $W'$
production~\cite{Papaefstathiou:2009sr}, Higgs boson production via
gluon fusion~\cite{Alioli:2008tz,Hamilton:2009za} and Higgs boson
production associated with a vector boson
(Higgs-strahlung)~\cite{Hamilton:2009za}. Unlike the \MCatNLO{}
implementation, \POWHEG{} produces events with positive (constant)
weight, and, furthermore, does not depend on the subsequent Shower
Monte Carlo program. It can be easily interfaced to any modern shower
generator and, in fact, it has been interfaced to
\HERWIG{}~\cite{Corcella:2000bw,Corcella:2002jc} and
\PYTHIA{}~\cite{Sjostrand:2006za} in
refs.~\cite{Nason:2006hfa,Frixione:2007nw,Alioli:2008gx,Alioli:2008tz}.

Single top production is the first \POWHEG{} implementation of a
process that has both initial- and final-state singularities,
and so the present work can serve as an example of how to deal
with this problem in \POWHEG{}.

This paper is organized as follows. In sec.~\ref{sec:description} we collect
the next-to-leading order cross section formulae and describe the kinematics
and the structure of the singularities.  In
sec.~\ref{sec:powheg_implementation} we discuss the {\POWHEG} implementation
and how we have included the generation of top-decay products.  In
sec.~\ref{sec:results} we show our results for several kinematic
variables. Most of this phenomenological section is devoted to study the
comparison of our results with those of \MCatNLO{}.  We find fair agreement
for almost all the distributions and give some explanations about the
differences we found. Some comparisons are carried out also with respect to
\PYTHIA{}~6.4, showing that some distributions are strongly affected by the
inclusion of NLO effects. Top-decay effects are also discussed.  Finally, in
sec.~\ref{sec:conc}, we give our conclusions.

\section{Description of the calculation}
\label{sec:description}
In this section we present some technical details of the calculation,
including the kinematic notation we are going to use throughout the paper,
the relevant differential cross sections up to next-to-leading order in the
strong coupling $\as$ and the subtraction formalism we have used to
regularize initial- and final-state singularities.  In this paper, we
always refer to top-quark production, since anti-top production is obtained
simply by charge conjugation.

Single-top production processes are usually divided into three
classes, depending on the virtuality of the $W$ boson involved at the
leading order:
\begin{enumerate}
\item
  Quark-antiquark annihilation processes, such as
\begin{equation}
\label{eq:s_ch_templ}
u + \bar{d} \to t + \bar{b} \,,
\end{equation}
are called $s$-channel processes since the $W$-boson virtuality is timelike.
\item
Processes where the top quark is produced with an exchange of a
spacelike $W$ boson, such as
\begin{equation}
\label{eq:t_ch_templ}
b + u \to t + d \, ,
\end{equation}
are called $t$-channel processes.
\item
Processes in which the top quark is produced in association with a real $W$
boson, such as
\begin{equation}
\label{eq:wt_ch_templ}
b + g \to t + W\, .
\end{equation}
These $Wt$ processes have a negligible cross section at the Tevatron, while
at the LHC their impact is phenomenologically relevant. The calculation of
NLO corrections to $Wt$ processes is also interesting from the theoretical
point of view, since the definition of real corrections is not
unambiguous~\cite{Frixione:2008yi}.
\end{enumerate}

In this paper we consider only $s$- and $t$-channel processes.
In these cases, the \POWHEG{} implementation needs to deal with both
initial- and final-state singularities, and is thus more involved
than in processes previously considered. The associated $Wt$ production
has only initial-state singularities and is thus analogous to
previous \POWHEG{} implementations. We leave it to a future work.

In the calculation, all quark masses have been set to zero (except, of
course, the top-quark mass) and the full Cabibbo-Kobayashi-Maskawa~(CKM)
matrix has been taken into account. However, for sake of illustration, we set
the CKM matrix equal to the identity in this section.

We refer to ref.~\cite{Frixione:2007vw} for the notation and for a deeper
description of the \POWHEG{} method. Here we just recall that with
$\mathcal{B}$, $\mathcal{V}$, $\mathcal{R}$ and $\mathcal{G}$ we indicate
the Born, virtual, real and collinear contributions respectively, divided by
the corresponding flux factor.  The same letters, capitalized, are used for
quantities multiplied by the luminosity factor. The explicit formulae
for these quantities are collected in sec.~\ref{sec:cross_sec}.

\subsection{Contributing subprocesses}
\label{sec:subprocesses}
In the following, we organize and label all the Born and real subprocesses,
keeping the distinction between the $s$- and $t$-channel contributions. This
distinction holds also when real corrections are considered, since, due to
color flow, interferences do not arise between real corrections to $s$- and
$t$-channel Born processes.
\begin{enumerate}[1)]
\item
In the $s$-channel case, there are only Born processes of the type $q q' \to
t \bar{b}$ , where $q$ and $q'$ run over all possible different quark and
antiquark flavours compatible with the final state.  We denote with
$\mathcal{B}_{qq'}$ the (summed and averaged) squared amplitude, divided by
the flux factor.  The corresponding real correction contributions include
processes with an outgoing or an incoming gluon, i.e.~processes of type $q q'
\to t \bar{b} g$, $g q \to t \bar{b} q'$ and $q g \to t \bar{b} q'$.  We
denote these contributions with $\mathcal{R}_{q q'}$, $\mathcal{R}_{gq, (s)}$
and $\mathcal{R}_{qg, (s)}$.  Summarizing, we have
\begin{center}
\begin{tabular}{|c|c|c|}
\hline
process & notation & contributing subprocesses \\
\hline
\hline
$q q' \to t \bar{b}$ & $\mathcal{B}_{qq'}$ & $u \bar{d} \to t
\bar{b}$, $\bar{d} u \to t \bar{b}$ \\ 
\hline
\hline
$q q' \to t \bar{b} g$ & $\mathcal{R}_{q q'}$ & $u \bar{d} \to t
\bar{b} g$, $\bar{d} u \to t \bar{b} g$ \\ 
\hline
$g q \to t \bar{b} q'$ & $\mathcal{R}_{gq, (s)}$ & $g u \to t \bar{b} d$, $g
\bar{d} \to t \bar{b} \bar{u}$ \\ 
\hline
$q g \to t \bar{b} q'$ & $\mathcal{R}_{qg, (s)}$ & $u g \to t \bar{b} d$,
$\bar{d} g \to t \bar{b} \bar{u}$ \\ 
\hline
\end{tabular}
\end{center}
where  $u$ and $d$ stand for a generic up- or down-type light quark. 

\item
In the $t$-channel case, there are only Born processes of the type $q
b \to t q'$ (and $b q \to t q'$), where $q$ and $q'$ run over all
possible flavours and anti-flavours. Their contributions are denoted
$\mathcal{B}_{q b}$ ($\mathcal{B}_{b q}$).  We use this notation since
we want to keep track of the down-type quark connected to the top
quark. The structure of real corrections is more complex in this case.
Contributions obtained from the previous processes by simply adding an
outgoing gluon, $q b \to t q' g$, will be denoted as $\mathcal{R}_{q
b}$.
The subprocesses generated by an initial-state gluon splitting into a 
quark-antiquark pair are designated by 
$\mathcal{R}_{q g, (t)}$ for $q g \to t q' \bar{b}$ {\big (}$\mathcal{R}_{g q, (t)}$
for $g q \to t q' \bar{b}${\big )} and $\mathcal{R}_{g b}$ for $g b \to t \bar{q} q'$
{\big (}$\mathcal{R}_{b g}$ for $b g \to t \bar{q} q'${\big )}.
In the former case $q$ and $q'$ are connected via a $Wqq'$ vertex,
while the gluon splits into a $b \bar{b}$ pair, so the top quark is
color connected with the incoming gluon.  In the latter case the
situation is opposite, since the gluon splits into a $q\bar{q}$ pair,
while the incoming $b$ is directly CKM-connected to the top quark. This
gives rise to a different singularities structure, which we take into
account in dealing with the $q g \to t q' \bar{b}$ ($g q \to t q'
\bar{b}$) and $g b \to t \bar{q} q'$ ($b g \to t \bar{q} q'$)
processes separately. 
Summarizing
\begin{center}
\begin{tabular}{|c|c|c|}
\hline
process & notation & contributing subprocesses \\
\hline
\hline
$b q \to t q'$ & $\mathcal{B}_{b q}$ & $b u \to t d$, $b \bar{d} \to t \bar{u}$ \\
\hline
$q b \to t q'$ & $\mathcal{B}_{q b}$ & $u b \to t d$, $\bar{d} b \to t \bar{u}$ \\
\hline
\hline
$b q \to t q' g$ & $\mathcal{R}_{b q}$ & $b u \to t d g$, $b \bar{d} \to t \bar{u} g$ \\
\hline
$q b \to t q' g$ & $\mathcal{R}_{q b}$ & $u b \to t d g$, $\bar{d} b \to t \bar{u} g$ \\
\hline
$g q \to t q' \bar{b}$ & $\mathcal{R}_{g q, (t)}$ & $g u \to t d \bar{b}$, $g \bar{d} \to t \bar{u} \bar{b}$ \\
\hline
$q g \to t q' \bar{b}$ & $\mathcal{R}_{q g, (t)}$ & $u g \to t d \bar{b}$, $\bar{d} g \to t \bar{u} \bar{b}$ \\
\hline
$g b \to t \bar{q} q'$ & $\mathcal{R}_{g b}$ & $g b \to t \bar{u} d$ \\
\hline
$b g \to t \bar{q} q'$ & $\mathcal{R}_{b g}$ & $b g \to t \bar{u} d$ \\
\hline
\end{tabular}
\end{center}
\end{enumerate}
where $u$ and $d$ stand for a generic up- or down-type light quark.  

\begin{figure}[thb] 
\centerline{ 
\subfigure[ $s$-channel]{
\epsfig{figure=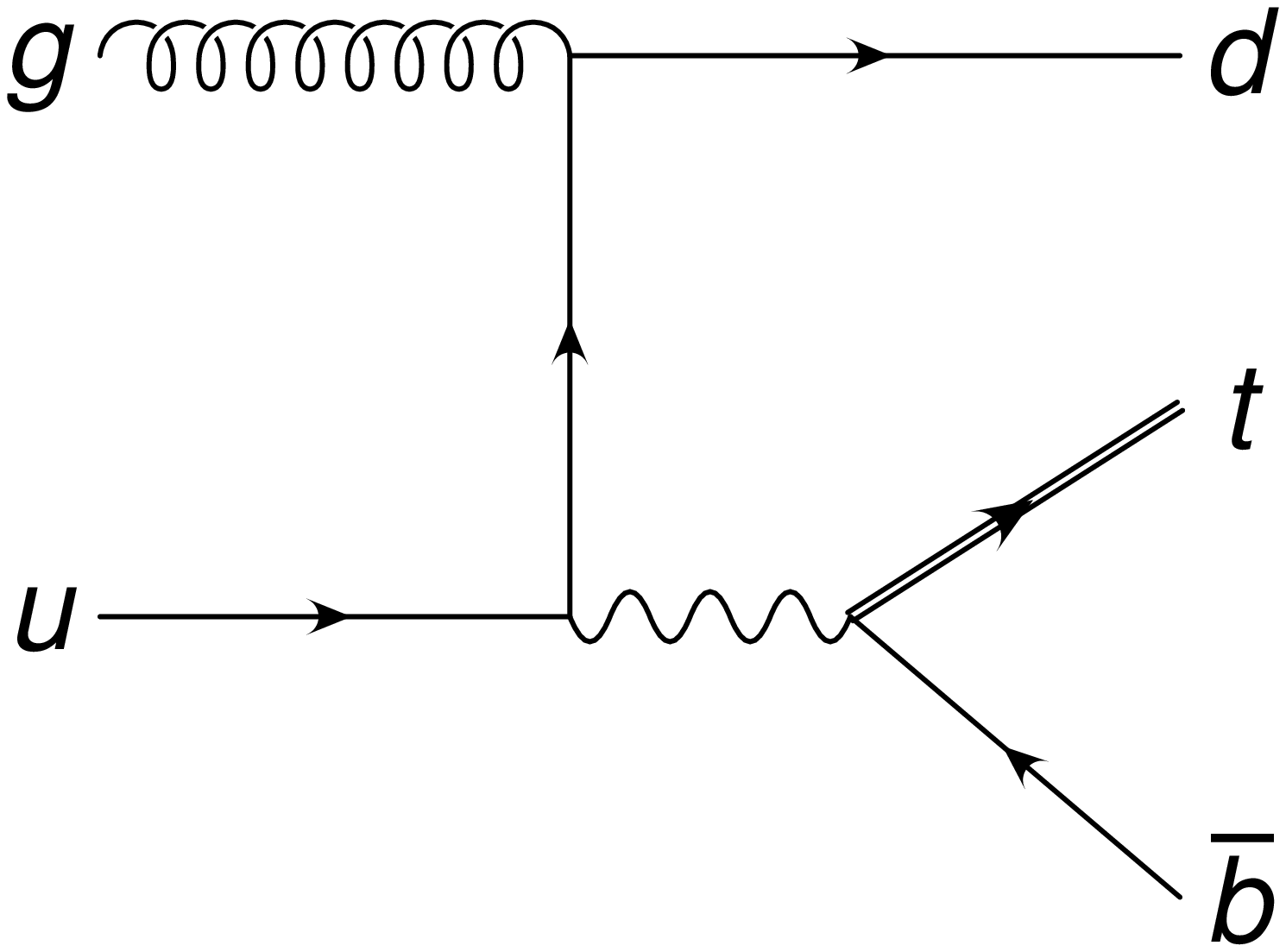,width=0.24\textwidth,clip=}  
\epsfig{figure=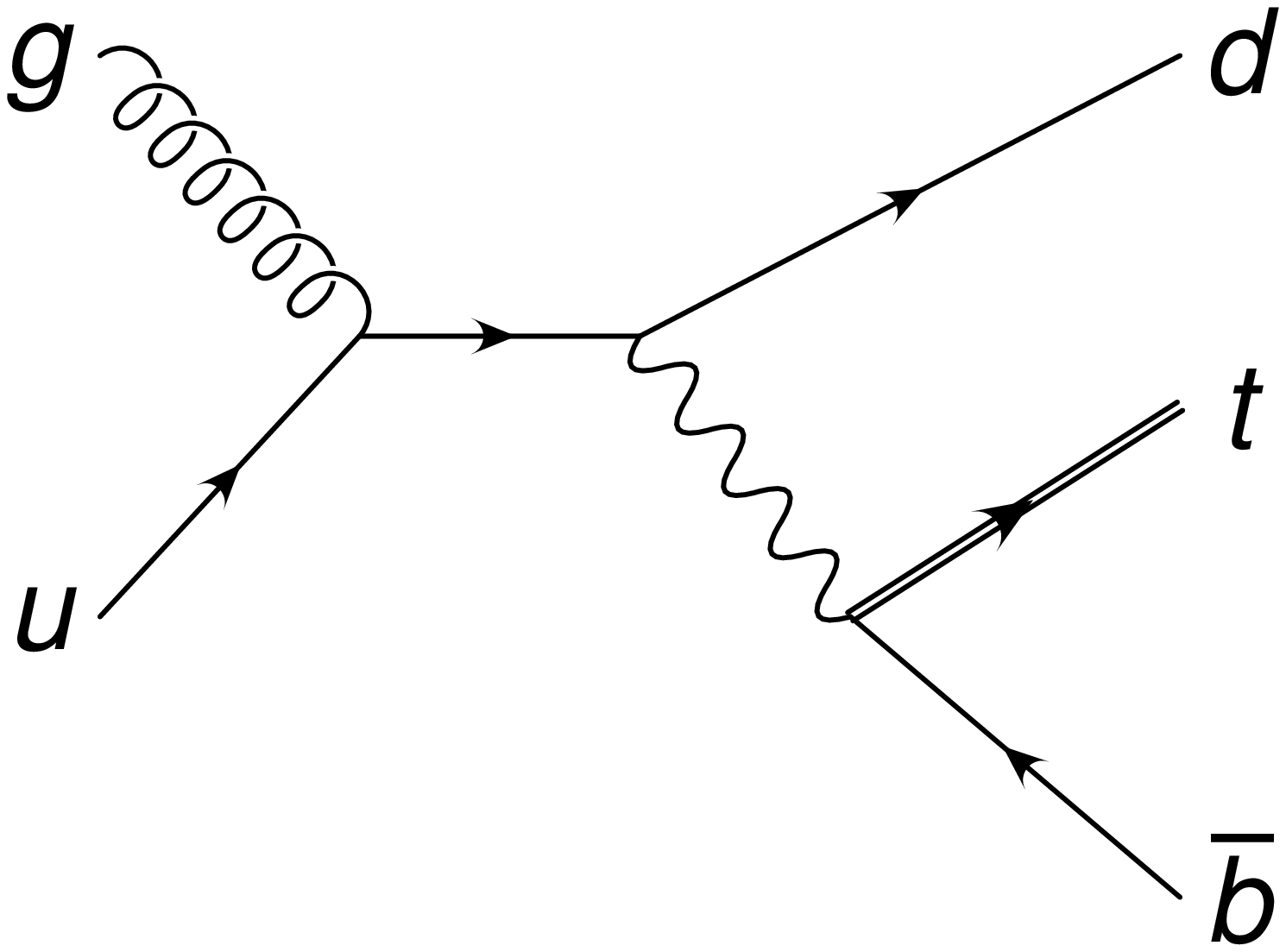,width=0.24\textwidth,clip=}}
\quad
\subfigure[ $t$-channel]{ 
\epsfig{figure=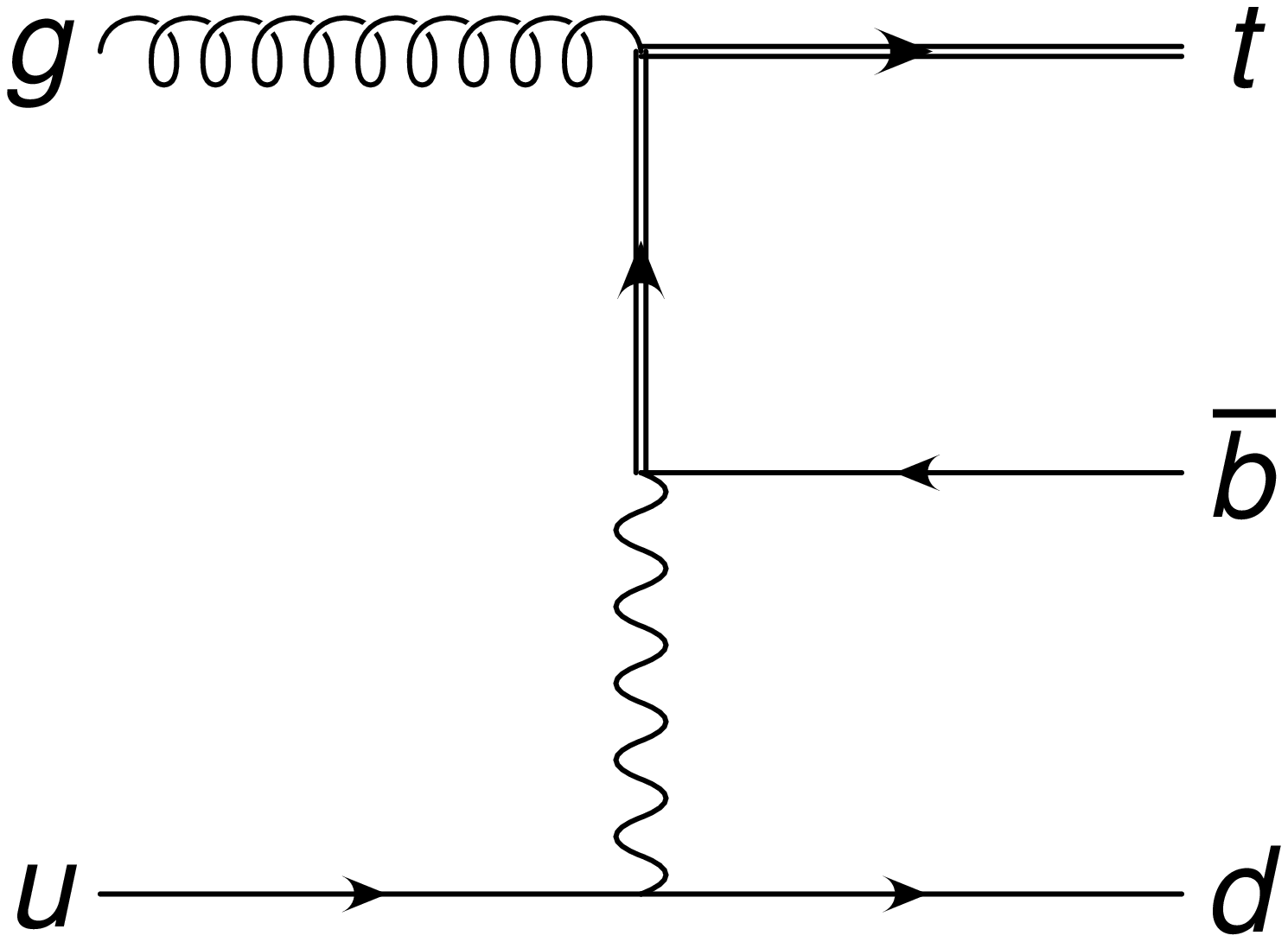,width=0.24\textwidth,clip=}
\epsfig{figure=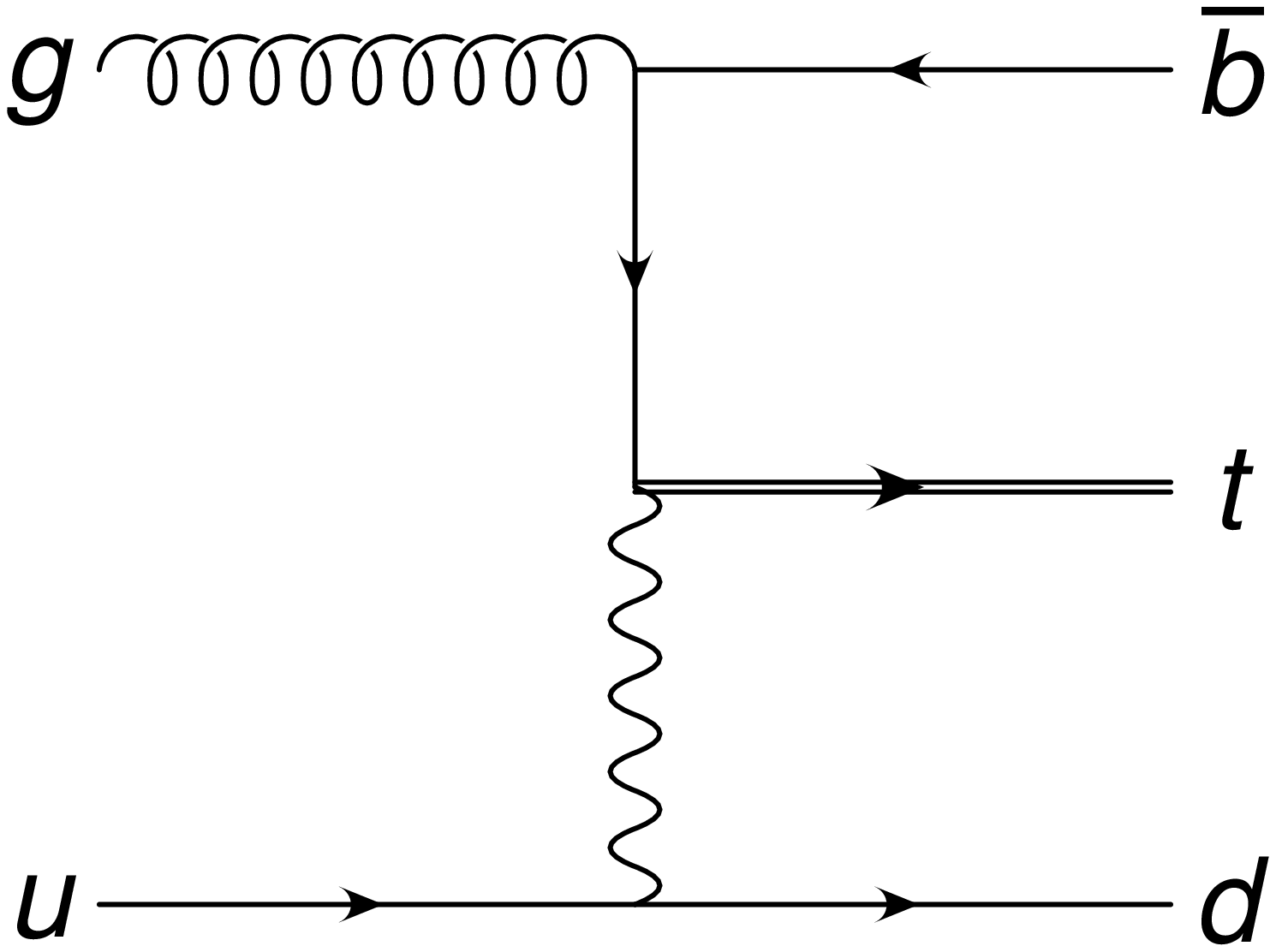,width=0.24\textwidth,clip=}}
} 
\caption{
Graphs corresponding to $s$- and $t$-channel contributions to the real
scattering $g u \to t d \bar{b}$.}
\label{fig:gu_tbd_st}
\end{figure}

In order to distinguish $s$- and $t$-channel real processes with the
same flavour structure, we have used the subscript~$(s)$ and~$(t)$ on
the $\mathcal{R}_{g q}$ and $\mathcal{R}_{q g}$ contributions.  As
already stated, these contributions do not interfere owing to
different color structures, so we can keep them distinct.  We have
drawn a sample of Feynman diagrams for $s$- and $t$-channel $g u \to t
d \bar{b}$ scattering in fig.~\ref{fig:gu_tbd_st}.  \\

\subsection{Kinematics and singularities structure}
\subsubsection{Born kinematics}
\label{sec:Born_kinematics}
Following the notation of ref.~\cite{Frixione:2007vw}, we denote with
$\bar{k}_{\splus}$ and $\bar{k}_{\sminus}$ the incoming quark
momenta, aligned along the plus and minus direction of the $z$ axis,
by $\bar{k}_1$ the outgoing top-quark momentum and by $\bar{k}_2$ the
other outgoing light-parton  momentum.  
The final-state top-quark virtuality will be denoted by $M^2$, so that
$\bar{k}_1^2=M^2$. The top quark on-shell condition is $M^2=m_t^2$, where
$m_t$ is the top-quark mass.
If $K_\splus$ and $K_\sminus$ are the momenta of the incoming hadrons,
then we have
\begin{equation}
\bar{k}_\splusminus=\bar{x}_\splusminus K_\splusminus \,,
\end{equation}
where $\bar{x}_\splusminus$ are the momentum fractions, and momentum
conservation reads
\beq
\bar{k}_\splus + \bar{k}_\sminus = \bar{k}_1 + \bar{k}_2\,.
\eeq
We introduce the variables
\begin{equation}
  \bar{s} = (\bar{k}_\splus+\bar{k}_\sminus)^2 ,\qquad\quad \bar{Y} =
  \frac{1}{2} 
  \log \frac{(\bar{k}_\splus + \bar{k}_\sminus)^0 + (\bar{k}_\splus +
  \bar{k}_\sminus)^3} {(\bar{k}_\splus + \bar{k}_\sminus)^0 -
  (\bar{k}_\splus + \bar{k}_\sminus)^3}\,,
\end{equation}
and $\bar{\theta}_1$, the angle between the outgoing top quark
and the $\bar{k}_{\splus}$ momentum, in the partonic center-of-mass~(CM)
frame. We denote with $\bar{\phi}_1$ the azimuthal angle of the
outgoing top quark in the same reference frame.
Since the differential cross sections do not depend on the overall azimuthal
orientation of the outgoing partons, we set this angle to zero.
At the end of the generation of an event, we perform a uniform,
random azimuthal rotation of the whole event, in order to cover the whole
final-state phase space.  The set of variables 
$\tmmathbf{\bar{\Phi}}_2 \equiv\lg \bar{s},\bar{Y},\bar{\theta}_1,
\bar{\phi}_1\rg$
fully parametrizes the Born kinematics. From them, we can reconstruct
the momentum fractions
\begin{equation}
\label{eq:xpxm_bar}  
\bar{x}_{\splus}  = \sqrt{\frac{\bar{s}}{S}}\, e^{\bar{Y}}, \qquad\quad
\bar{x}_{\sminus} = \sqrt{\frac{\bar{s}}{S}}\, e^{- \bar{Y}}\,,
\end{equation}
where $S=(\Kplus + \Kminus)^2$ is the squared CM energy of
the hadronic collider.  The outgoing momenta are first reconstructed
in their longitudinal rest frame, where $\bar{Y}=0$. In this frame, their
energies are
\begin{equation}
\bar{k}_1^0 |_{\sss \bar{Y}=0}=\sqrt{\(\frac{\bar{s}-M^2}{2\sqrt{\bar{s}}}\)^2+M^2}
\qquad\mbox{and}\qquad
\bar{k}_2^0 |_{\sss \bar{Y}=0}=\frac{\bar{s}-M^2}{2\sqrt{\bar{s}}}.
\end{equation}
The two spatial momenta are obviously opposite and both have
modulus equal to $\bar{k}_2^0|_{\sss \bar{Y}=0}$. We fix the top-quark momentum
to form an angle $\bar{\theta}_1$ with the $\splus$ direction and to have
zero azimuth (i.e.~it lies in the $x z$ plane and has positive $x$
component). Both $\bar{k}_1$ and $\bar{k}_2$ are then boosted back in
the laboratory frame, with  boost rapidity $\bar{Y}$.  The Born phase
space, in terms of these variables, can be written as
\begin{eqnarray}
\label{eq:Bphsp}
d \tmmathbf{\bar{\Phi}}_2 &=& d\bar{x}_\splus \, d\bar{x}_\sminus
(2\pi)^4 \delta^4\!\(\bar{k}_\splus+ \bar{k}_\sminus -\bar{k}_1 -\bar{k}_2\)
\frac{d^3 \bar{k}_1}{(2\pi)^3 2 \bar{k}_1^0} \,
\frac{d^3 \bar{k}_2}{(2\pi)^3 2 \bar{k}_2^0} \nonumber\\
&=&\frac{1}{S}\frac{\beta}{16\pi}\ d\bar{s}\ d\bar{Y}\ d\cos{\bar{\theta}_1}\
\frac{d\bar{\phi}_1}{2\pi}\,,
\end{eqnarray}
where
\begin{equation}
\beta=1-\frac{M^2}{\bar{s}}\, .
\end{equation}
We generate the top quark with virtuality
$M^2$ and decay it with a method analogous to the one adopted
in ref.~\cite{Frixione:2007zp}, that will be described in
sec.~\ref{sec:top_decay}. We take into account the top finite width
by first introducing a trivial integration $\int dM^2\ \delta(M^2-m_t^2)$ in
eq.~(\ref{eq:Bphsp}) and then by performing the replacement
\begin{equation}
\label{eq:breitwigner}
\delta\!\(M^2 -m_t^2\) \, \to \, \frac{1}{\pi} \frac{m_t\,\Gamma_t}
{\(M^2 - m_t^2\)^2 + ( m_t^2\, \Gamma_t^2 ) }\,.
\end{equation}
With this substitution, the final expression for the Born phase space
reads
\begin{equation}
d \tmmathbf{\bar{\Phi}}_2 = \frac{1}{S}\frac{\beta}{16\pi^2}\,
\frac{m_t\,\Gamma_t}{\(M^2 - m_t^2\)^2 +  m_t^2\, \Gamma_t^2  }
\, d M^2\,  d\bar{s}\, d \bar{Y} \, d\cos{\bar{\theta}_1}\,
\frac{d\bar{\phi}_1}{2\pi}\,.
\end{equation} 

\subsubsection{Real-emission kinematics}
\label{sec:real_kinematics}
Real-emission processes have an additional final-state parton that can be
emitted from an incoming leg only ($\mathcal{R}_{gq, (s)}$, $\mathcal{R}_{qg,
  (s)}$, $\mathcal{R}_{g q, (t)}$, $\mathcal{R}_{q g, (t)}$, $\mathcal{R}_{g
  b}$, $\mathcal{R}_{b g}$) or from both an initial- and final-state leg
($\mathcal{R}_{q q'}$, $\mathcal{R}_{b q}$, $\mathcal{R}_{q b}$).  We
need then to use two different parametrizations of the real phase space, one
to deal with initial-state singularities and one for final-state ones. We
treat the radiation kinematics according to the variant of the \FKS{}~(FKS)
subtraction scheme~\cite{Frixione:1995ms,Frixione:1997np} illustrated
in ref.~\cite{Frixione:2007vw}. Before giving all the technical details, we
summarize briefly how the procedure works:
\begin{itemize}
\item[-] We split each real squared amplitude into contributions
that have at most one collinear (and/or one soft) singularity.
\item[-] We build the collinear (and soft) subtraction
terms needed to deal with that singularity.
\item[-] We choose the emission phase-space parametrization
suited for the singularity we integrate on.
\end{itemize}
In the FKS~method, the singular regions associated with the 
$\splus$ and $\sminus$ legs are treated with the same
kinematics. Nevertheless, we have decided to split these two different
contributions in order to gain a clear subtraction structure.

We now describe the procedure used to split real
squared amplitudes and the corresponding phase-space parametrizations.
Subtraction terms will then be given in sec.~\ref{sec:softcoll_limits}.
We proceed as follows:
\begin{enumerate}
\item
We start by considering real processes that have both initial- and
final-state emissions, namely the $\mathcal{R}_{q q'}$,
$\mathcal{R}_{b q}$ and $\mathcal{R}_{q b}$ contributions.
In this case, the FKS parton is the outgoing gluon and we
choose it to be the last particle. We denote its momentum
by $k_3$, so that momentum conservation reads
\begin{equation}
\label{eq:real_mom_cons}
k_\splus + k_\sminus = k_1 + k_2 + k_3,
\end{equation}
where $k_\splus$, $k_\sminus$, $k_1$ and $k_2$ label the same particles of
the underlying Born process.  The FKS parton can become collinear to one of
the incoming legs or to the other massless final-state leg, so we need to
introduce a set of functions to project out these different singular
regions. The general properties these functions have to satisfy were given
in sec.~2.4 of ref.~\cite{Frixione:2007vw}.  In this paper we  use
\begin{equation}
\mathcal{S}^{3,\splus} = {\cal D}^{-1}\ \frac{1}{d_{3,\splus}}\,, \qquad
\mathcal{S}^{3,\sminus} = {\cal D}^{-1}\ \frac{1}{d_{3,\sminus}}\,, \qquad
\mathcal{S}^{3,2} = {\cal D}^{-1}\ \frac{1}{d_{3,2}}\,,
\end{equation}
where
\begin{equation}
{\cal D}= {\frac{1}{d_{3,\splus}} + \frac{1}{d_{3,\sminus}} + \frac{1}{d_{3,2}}}
\qquad\mbox{and}\qquad d_{i,j}=k_i \cdot k_j\,.
\end{equation}
For any given kinematic configuration, these functions satisfy
\begin{equation}
\mathcal{S}^{3,\splus} + \mathcal{S}^{3,\sminus} + \mathcal{S}^{3,2} = 1.
\end{equation}
The separation among different singular regions is performed
multiplying each real contribution with the corresponding $\mathcal{S}$
function. For example, for the $s$-channel $\mathcal{R}_{q q'}$ case, we
have 
\begin{eqnarray}
\mathcal{R}_{q q'}^{3,\splus} &=& \mathcal{R}_{q q'}
     \ \mathcal{S}^{3,\splus}, \nonumber \\
\mathcal{R}_{q q'}^{3,\sminus} &=& \mathcal{R}_{q q'}
     \ \mathcal{S}^{3,\sminus}, \nonumber \\
\mathcal{R}_{q q'}^{3,2} &=& \mathcal{R}_{q q'}
     \ \mathcal{S}^{3,2}. 
\end{eqnarray}
These contributions are now singular only when the FKS parton becomes
collinear to $k_\splus$, $k_\sminus$ and $k_2$ respectively, or soft.
Analogous relations hold for $\mathcal{R}_{b q}$ and
$\mathcal{R}_{q b}$.

\item
Next we consider the real process $g b \to t \bar{q} q'$. It is
singular when $\bar{q}$ or $q'$ become collinear to the incoming
gluon, so that the FKS parton can be respectively $\bar{q}$ or $q'$
and we need again a set of functions to project out the different
singular regions.  Recalling the labeling of the momenta
\begin{equation}
g\,(k_\splus)\,b\,(k_\sminus) \to\,t\,(k_1)\,\bar{q}\,(k_2)\,q'\,(k_3)\,,
\nonumber
\end{equation}
we introduce the projecting functions
\begin{eqnarray}
\label{eq:S_gb}
\mathcal{S}^{2,\splus} &=&
\(\frac{1}{d_{2,\splus}} + \frac{1}{d_{3,\splus}}\)^{-1}
\frac{1}{d_{2,\splus}}\,, \nonumber\\
\mathcal{S}^{3,\splus} &=&
\(\frac{1}{d_{2,\splus}} + \frac{1}{d_{3,\splus}}\)^{-1}
\frac{1}{d_{3,\splus}}\,,
\end{eqnarray}
to isolate the region where $k_2 \cdot k_\splus \to 0$ or $k_3 \cdot
k_\splus \to 0$. 
We have then the two contributions
\begin{eqnarray}
\label{eq:R_gb}
\mathcal{R}_{g b}^{3,\splus} &=& \mathcal{R}_{g b} \ \mathcal{S}^{3,\splus}\,,
\nonumber \\ 
\mathcal{R}_{g b}^{2,\splus} &=& \mathcal{R}_{g b} \ \mathcal{S}^{2,\splus}\,, 
\end{eqnarray}
coming from $\mathcal{R}_{g b}$. For $b g \to t \bar{q} q'$, analogous
contributions can be obtained from eqs.~(\ref{eq:S_gb}) and
(\ref{eq:R_gb}) with the substitutions $\mathcal{R}_{gb} \to
\mathcal{R}_{bg}$ and $\splus \to \sminus$.
\item
To deal with the remaining real contributions we do not need to introduce
any other $\mathcal{S}$ function, since each of them is singular in  one
region only (the $\splus$ one for $\mathcal{R}_{g q, (s)}$ and
$\mathcal{R}_{g q, (t)}$, the $\sminus$ one for $\mathcal{R}_{q g,
(s)}$ and $\mathcal{R}_{q g, (t)}$).
\end{enumerate}
Having split all real contributions in such a way that each term has at
most one singularity, we can associate with each of them a particular
phase-space parametrization, suitable to handle that singularity structure.  
In the following we summarize the reconstruction procedure needed to build
the real-emission kinematics, given the underlying Born one, and a set of
three radiation variables.  For all the details, we refer to sec.~5 of
ref.~\cite{Frixione:2007vw}.

\subsubsection*{Parametrization of the initial-state radiation~(ISR) phase
  space} 
The FKS method uses the same phase-space parametrization for
describing both the~$\splus$ and~$\sminus$ singular regions. The set
of radiation variables
\begin{equation}
\label{eq:FKSradvars}
\ISRRad=\{ \xi,\ y,\ \phi\}\,,
\end{equation}
together with the Born ones, completely reconstruct the real-event
kinematics: $\tmmathbf{\Phi}_3 \equiv \lg \bar{s}, \bar{Y},
\bar{\theta}_1, \xi, y, \phi\rg$.  Using eq.~(\ref{eq:xpxm_bar}), we
can compute the underlying Born momentum fractions $\bxplusminus$ and,
from them, we obtain
\begin{equation}
\label{eq:xp_xm_IS_FKS}
 x_\splus = \frac{\bar{x}_\splus}{ \sqrt{1-\xi}} \sqrt{\frac{2-\xi
      (1-y)}{2-\xi (1+y)}}, \qquad\qquad
 x_\sminus= \frac{\bar{x}_\sminus}{ \sqrt{1-\xi}} \sqrt{\frac{2-\xi
      (1+y)}{2-\xi(1-y)}},
\end{equation}
with the kinematics constraints 
\begin{equation}
0 \le \xi \le \ximax(y) \,,
\end{equation}
where 
\begin{eqnarray}
\label{eq:ximax}
 \ximax(y)= 1- {\rm max}\!\!\!&&\lg
\frac{2(1+y)\,\bar{x}_\splus^2}{\sqrt{(1+\bar{x}_\splus^2)^2(1-y)^2 +
 16\,y\,\bar{x}_\splus^2}+(1-y)(1-\bar{x}_\splus^2)},\right.
\nonumber\\
&&\phantom{\Bigg\{}\left.
\frac{2(1-y)\,\bar{x}_\sminus ^2}{\sqrt{(1+\bar{x}_\sminus ^2)^2(1+y)^2 -
    16\,y\,\bar{x}_\sminus ^2} +(1+y)(1-\bar{x}_\sminus ^2)}\rg .
\end{eqnarray}
In the laboratory frame, the incoming momenta are given by
\begin{equation}
k_\splusminus=x_\splusminus K_\splusminus \,.
\end{equation}
In the partonic center-of-mass frame, we define the FKS parton to
have momentum
\begin{equation}
\label{eq:FKS_mom_ISR}
  k'_3 = k_3^{\prime\, 0}\ (1, \sin\theta \sin\phi, \sin\theta \cos\phi,
  \cos\theta), 
\end{equation}
where
\begin{equation}
\label{eq:FKS_csi_ISR}
  k_3^{\prime\, 0} = \frac{\sqrt{s}}{2} \xi,
  \qquad \cos\theta = y\,,
\end{equation} and 
\begin{equation}
s=\(k_\splus+k_\sminus\)^2=\frac{\bar{s}}{1-\xi}.
\end{equation}
From eqs.~(\ref{eq:FKS_mom_ISR}) and~(\ref{eq:FKS_csi_ISR}), we see that the
soft limit is approached when $\xi\to 0$, while the collinear limits are
characterized by $y\to 1$ ($k_3$ parallel to the $\splus$ direction) or $y\to
-1$ ($k_3$ parallel to the $\sminus$ direction).

Boosting $k'_3$ back in the laboratory frame with longitudinal velocity
$(\xplus - \xminus)/(\xplus + \xminus)$ we obtain $k_3$.  Having computed
$k_3$ and $k_\splusminus$, we can construct $k_{\rm tot}=k_\splus +k_\sminus
- k_3$, while from the underlying Born momenta we have $\bar{k}_{\rm
  tot}=\bar{k}_1 + \bar{k}_2$.
We construct then the longitudinal boost $\boost_L$, with boost velocity
$\vec{\beta}_L=(0,0,\beta_L)$, where
\begin{equation}
\beta_L = - \frac{\bxplus - \bxminus}{\bxplus + \bxminus}\,,
\end{equation}
so that the boosted momentum  $k''_{\rm tot}=\boost_L k_{\rm tot}$ has zero
longitudinal component.
In addition we define 
\begin{equation}
\vec{\beta}_T=-\frac{\vec{k}''_{\rm tot}}{k^{\prime\prime\,0}_{\rm tot}}
\end{equation}
and the corresponding (transverse) boost $\boost_T$, so that
$\boost_T k''_{\rm tot}$ has zero transverse momentum.
The final-state momenta $k_1$ and $k_2$ in the laboratory frame are obtained
with the following boost sequence
\begin{equation}
k_i =  \boost_L^{-1}\, \boost_T^{-1} \, \boost_L \,
\bar{k}_i\,, \qquad  i=1,2\,.
\end{equation}
Finally, the three-body phase space can  be written, in a factorized form, in
terms of the Born and radiation phase space
\beq
d \tmmathbf{\Phi}_3 = dx_\splus \, dx_\sminus (2\pi)^4 \delta^4
(k_\splus+ k_\sminus -k_1-k_2-k_3) \frac{d^3 k_1}{(2\pi)^3 2 k_1^0}
\frac{d^3 k_2}{(2\pi)^3 2 k_2^0} \frac{d^3 k_3}{(2\pi)^3 2 k_3^0}
= d\tmmathbf{\bar{\Phi}}_2\,d\ISRRad  \,,
\eeq
where
\beq
\label{eq:dRad_IS_FKS}
d\ISRRad = \frac{s}{(4\pi)^3}\,\frac{\xi}{1-\xi}\,d\xi\,dy\, d\phi \equiv
J_{\rm rad}^{\sss\rm ISR}\!\(\tmmathbf{\bar{\Phi}}_2,\ISRRad\)\,d\xi\,dy\,
d\phi\,,
\eeq
that defines the Jacobian $J_{\rm rad}^{\sss\rm ISR}$ of the change of
variables. 

\subsubsection*{Parametrization of the final-state radiation~(FSR) phase
  space}
For the FSR phase-space parametrization $\FSRRad$, we use the same notation
as for the initial-state case $\ISRRad$ (see eq.~(\ref{eq:FKSradvars})). We
define, in the partonic center-of-mass frame,
\begin{equation}
\label{eq:FKSfsr}
\xi=\frac{2 k_{3}^0}{q^0}\,,\qquad \quad
y=\frac{\vec{k}_{3}\cdot \vec{k}_{2}}{\mmod{k}_{3}\ \mmod{k}_2}\,,\qquad
\quad 
\phi=\phi\(\vec{\eta} \times \vec{k},\; \vec{k}_{3} \times \vec{k}\),
\end{equation}
where
\begin{equation}
\label{eq:FKS_q_k}
q=k_\splus + k_\sminus\,, \qquad k = k_2 + k_3\,,
\end{equation}
and the notation $\mmod{p}$ stands for $|\vec{p}|$.  We denote with
$\vec\eta\,$ an arbitrary direction that serves as origin for the azimuthal
angle of $\vec{k}_{3}$ around $\vec{k}$, while ``$\times$'' is the cross
vector product.  The notation $\phi(\vec{v}_1,\vec{v}_2)$ indicates the angle
between $\vec{v}_1$ and $\vec{v}_2$, so that $\phi$ is the azimuth of the
vector $\vec{k}_{3}$ around the direction $\vec{k}$.\footnote{The FKS variant
  that we use (see ref.~\cite{Frixione:2007vw}) has a slightly different
  definition of $\phi$ than the one introduced in the original FKS papers.}

From eq.~(\ref{eq:FKSfsr}) we see that the soft limit is approached when
$\xi\to 0$, while the collinear limit is characterized by $y\to 1$ ($k_3$
parallel to $k_2$).

Given the set of variables $\tmmathbf{\Phi}_3 \equiv \lg \bar{s}, \bar{Y},
\bar{\theta}_1, \xi, y, \phi\rg$ we can reconstruct the full real-event
kinematics.  The momentum fractions $\xplusminus$ are the same as the
underlying Born ones, since the emission from a final-state leg does not
affect them, so that
\begin{equation}
\xplus=\bar{x}_\splus\,, \qquad\quad \xminus=\bar{x}_\sminus\, 
\qquad\quad\mbox{and} \qquad\quad s=\bar{s}\,.
\end{equation}
Inverting the first identity in eq.~(\ref{eq:FKSfsr}), we immediately have
\begin{equation}
k_{3}^0=\mmod{k}_3=\xi\frac{q^0}{2}\,,
\end{equation}
where $\xi$ is limited by
\begin{equation}
0\le\xi\le\ximax \equiv \frac{q^2-M_{\rm rec}^2}{q^2}\,,
\end{equation}
with
\begin{equation}
M_{\rm rec}^2 = (q-\bar{k}_2)^2 = k_1^2\,.
\end{equation}
The energy (and the modulus) of the other light outgoing parton, always in
the partonic center-of-mass frame, is given by
\begin{equation}
\label{eq:inversefkskn}
k_2^0=\mmod{k}_2=\frac{q^2-\Mrec^2-2q^0\mmod{k}_{3}}
{2\lq q^0-\mmod{k}_{3}\,(1-y)\rq}\;.
\end{equation}
Given $\mmod{k}_2$ and $\mmod{k}_3$ we construct the
corresponding vectors $\vec{k}_{2}$ and $\vec{k}_{3}$ such
that their vector sum $\vec{k}$ is parallel to $\vec{\bar{k}}_2$ and
the azimuth of $\vec{k}_{3}$ relative to $\vec{k}$ (the given
reference direction) is $\phi$.  Having fully defined $k_2$ and
$k_{3}$, we can reconstruct the vector $k$ of eq.~(\ref{eq:FKS_q_k})
and find $k_{\rm rec}=q-k$.
Finally, $k_1$ can be obtained boosting $\bar{k}_1$ along the $k_{\rm rec}$
direction with boost velocity
\begin{equation}
\vec{\beta}= - \(\frac{q^2-(k^0_{\rm rec}+\mmod{k}_{\rm rec})^2}
{q^2+(k^0_{\rm rec}+\mmod{k}_{\rm rec})^2}\) \frac{\vec{k}_{\rm
rec}}{\mmod{k}_{\rm rec}}\,,
\end{equation}
or, alternatively, exploiting momentum conservation of
eq.~(\ref{eq:real_mom_cons}).  To obtain the momenta in the laboratory frame
we need to boost back all the outgoing momenta computed in the center-of-mass
frame.

In this case too, the three-body phase space can be written in a
factorized form in terms of the Born and radiation phase space
\beq
d \tmmathbf{\Phi}_3 = dx_\splus \, dx_\sminus (2\pi)^4 \delta^4
(k_\splus+ k_\sminus -k_1-k_2-k_3) \frac{d^3 k_1}{(2\pi)^3 2 k_1^0}
\frac{d^3 k_2}{(2\pi)^3 2 k_2^0} \frac{d^3 k_3}{(2\pi)^3 2 k_3^0}
= d\tmmathbf{\bar{\Phi}}_2\,d\FSRRad  \,,
\eeq
where
\begin{eqnarray}
\label{eq:dRad_FS_FKS}
d\FSRRad &=&
\frac{q^2\,\xi}{(4\pi)^3}\,\frac{\mmod{k}_2^2}{\mmod{\bar{k}}_2}
\( \mmod{k}_2-\frac{k^2}{2q^0}\)^{-1} d\xi \,dy\, d\phi \nonumber \\
&=& \frac{s}{(4\pi)^3}\,\frac{4\,\xi}{\lq2-\xi\(1-y\)\rq^2}
\( 1-\frac{s\,\xi}{s-\Mrec^2}\) d\xi \,dy\, d\phi
\equiv J_{\rm rad}^{\sss\rm FSR}\(\tmmathbf{\bar{\Phi}}_2,\FSRRad\)\,d\xi\,dy\,
d\phi\,. 
\nonumber\\
\end{eqnarray}

\subsection{Squared amplitudes}
\label{sec:cross_sec}
In order to apply the \POWHEG{} method, we need the Born, real and
soft-virtual contributions to the differential cross section, i.e.~the
squared amplitudes, summed (averaged) over colors and helicities of
the outgoing (incoming) partons, and multiplied by the appropriate
flux factor. We have taken the Born, real and soft-virtual contributions
from the \MCatNLO{} code, testing, where possible, our
implementation against MadGraph subroutines~\cite{Alwall:2007st}.  All
the matrix elements have been evaluated in the zero-width
approximation, i.e.~$\Gamma_t$ and $\Gamma_W$ are set equal to zero in
all the propagators.  As already mentioned, to recover finite-width
effects in top-decay, the top mass $M$ is generated
according to  a Breit-Wigner distribution, centered in $m_t$ and with width
$\Gamma_t$ (see eq.~(\ref{eq:breitwigner})).

In the following, we give explicit expressions for the Born and
collinear remnant contributions. Real and soft-virtual matrix elements are
more complicated,
and we do not report them explicitly.  Nevertheless, we give the soft and
collinear limits of the real amplitude, since these expressions are needed
in the FKS subtraction formalism. 

\subsubsection{Born and virtual contributions}
\label{sec:Born}
We denote the $s$-channel squared matrix element for the lowest-order
contribution, averaged over color and helicities of the incoming particles,
and multiplied by the flux factor $1/(2\bar{s})$, as
$\mathcal{B}_{qq'}$.  For example, for the $u \bar{d} \to t \bar{b}$
subprocess, we have
\begin{equation}
\label{eq:b_ud}
\mathcal{B}_{u \bar{d}}=\frac{1}{2\bar{s}}\,\frac{g^4}{4}\bar{u}(\bar{u}-M^2)
\abs{\frac{1}{\bar{s}-m_W^2}}^2 |V_{ud}|^2|V_{tb}|^2,
\end{equation}
where $\bar{u}=(\bar{k}_{\splus}-\bar{k}_2)^2$ is the usual Mandelstam
variable, $g$ is the weak coupling ($e=g \sin\theta_W^{\rm eff}$) and
$V_{ij}$'s are the CKM matrix elements.
Crossing eq.~(\ref{eq:b_ud}) we have,  for the $\bar{d}u$ initiated process,
\begin{equation}
\label{eq:b_du}
\mathcal{B}_{\bar{d} u}=\frac{1}{2\bar{s}}\,\frac{g^4}{4}\bar{t}(\bar{t}-M^2)
\abs{\frac{1}{\bar{s}-m_W^2}}^2 |V_{ud}|^2|V_{tb}|^2 ,
\end{equation}
and for the $t$-channel contributions ($\mathcal{B}_{b q}$ and
$\mathcal{B}_{q b}$) of the $b u \to t d$ and $u b \to t d$ subprocesses
\begin{eqnarray}
\label{eq:b_bu_ub}
\mathcal{B}_{b u}&=&\frac{1}{2\bar{s}}\,\frac{g^4}{4}\bar{s}(\bar{s}-M^2)
\abs{\frac{1}{\bar{t}-m_W^2}}^2 |V_{ud}|^2|V_{tb}|^2, \nonumber\\
\mathcal{B}_{u b}&=&\frac{1}{2\bar{s}}\,\frac{g^4}{4}\bar{s}(\bar{s}-M^2)
\abs{\frac{1}{\bar{u}-m_W^2}}^2 |V_{ud}|^2|V_{tb}|^2 ,
\end{eqnarray}
where $\bar{t}=(\bar{k}_{\splus}-\bar{k}_1)^2$.
The corresponding expressions for $b \bar{d} \to t \bar{u}$
and $\bar{d} b \to t \bar{u}$ can be obtained from the latter
again by crossing. They are given by
\begin{eqnarray}
\label{eq:b_bd_db}
\mathcal{B}_{b \bar{d}}=\frac{1}{2\bar{s}}\,\frac{g^4}{4}\bar{u}(\bar{u}-M^2)
\abs{\frac{1}{\bar{t}-m_W^2}}^2 |V_{ud}|^2|V_{tb}|^2, \nonumber\\
\mathcal{B}_{\bar{d} b}=\frac{1}{2\bar{s}}\,\frac{g^4}{4}\bar{t}(\bar{t}-M^2)
\abs{\frac{1}{\bar{u}-m_W^2}}^2 |V_{ud}|^2|V_{tb}|^2.
\end{eqnarray}
The finite soft-virtual contributions, obtained according to the FKS method,
have been taken from the \MCatNLO{} code.  We included them in our NLO
calculation and tested the correct behaviour of our program by comparing our
NLO results with the MCFM code~\cite{MCFM}, both for the full NLO cross
section and for typical differential distributions. Some comparisons have
also been carried out with the program ZTOP~\cite{ZTOP}.

\subsubsection{Collinear remnants}
\label{sec:collinear}
The collinear remnants are given in eq.~(2.102)
of ref.~\cite{Frixione:2007vw}. Here we limit ourselves to list all the
contributions, giving only a couple of explicit examples to clarify the
notation.\\
For the $s$-channel processes, the collinear remnants are
\begin{equation}
\label{eq:sch_coll}
\mathcal{G}_{\splusminus}^{q q'}({\bf\Phi}_{2,\splusminus})\,, \qquad
\mathcal{G}_{\splus}^{gq}({\bf\Phi}_{2,\splus})\, \qquad\mbox{and}\qquad
\mathcal{G}_{\sminus}^{qg}({\bf\Phi}_{2,\sminus})\,,
\end{equation}
where the ${\bf \Phi}_{2,\splus}$ notation, according to
ref.~\cite{Frixione:2007vw}, represents the set of variables
\begin{equation}
\label{eq:collpluskin}
{\bf \Phi}_{2,\splus}=\{x_\splus,x_\sminus,z,k_1,k_2\},\qquad
{\rm with } \qquad
z\, x_\splus K_\splus +x_\sminus K_\sminus=k_1+k_2\,.
\end{equation}
The underlying Born configuration $\bar{\bf \Phi}_2$, associated
with the ${\bf \Phi}_{2,\splus}$ kinematics, is defined by
\begin{equation}
\bar{k}_\splus=z \,x_\splus K_\splus,\qquad \bar{k}_\sminus=x_\sminus
K_\sminus, \qquad \bar{k}_1=k_1,\quad \bar{k}_2=k_2\,.
\end{equation}
Similar formulae hold for ${\bf \Phi}_{2,\sminus}$.  Among the
contributions listed in~(\ref{eq:sch_coll}), only the real process
$q q'\to t\bar{b}g$ is singular in both the $\splus$ and the
$\sminus$ region. It thus needs the two collinear remnants
\begin{eqnarray}
\label{eq:sch_qq_collplusminus}
\mathcal{G}_{\splusminus}^{q q'}({\bf \Phi}_{2,\splusminus}) &=& \frac{\as}{2\pi}\CF
\Bigg\{ (1+z^2)\lq\(\frac{1}{1-z}\)_+
\log\frac{\bar{s}}{z\muF^2} +  2\(\frac{\log(1-z)}{1-z}\)_+\rq  \nonumber\\ && 
+ \ (1-z)  \Bigg\} \  {\cal B}_{q q'}(\bar{s},\bar{Y},\bar{\theta}_1)\,.
\end{eqnarray}
\\
For the $t$-channel processes, the collinear remnants are
\begin{equation}
\mathcal{G}_{\splusminus}^{bq}({\bf \Phi}_{2,\splusminus})\,,\quad
\mathcal{G}_{\splusminus}^{qb}({\bf \Phi}_{2,\splusminus})\,,\quad
\mathcal{G}_{\splus}^{gq}({\bf \Phi}_{2,\splus})\,,\quad
\mathcal{G}_{\sminus}^{qg}({\bf \Phi}_{2,\sminus})\,,\quad
\mathcal{G}_{\splus}^{gb}({\bf \Phi}_{2,\splus})\,\quad\mbox{and}\quad
\mathcal{G}_{\sminus}^{bg}({\bf \Phi}_{2,\sminus})\,.
\end{equation}
In this case, $\mathcal{G}_{\splus}^{gb}({\bf \Phi}_{2,\splus})$
contains two terms, since in the scattering $g b \to t \bar{q} q'$
both the two outgoing massless partons $\qb$ and $q'$ can become collinear
to the incoming gluon. We have
\begin{eqnarray}
\label{eq:tch_gb_collplus}
\mathcal{G}_{\splus}^{gb}({\bf \Phi}_{2,\splus}) &=& \frac{\as}{2\pi}\TF
\Bigg\{ (1-z) \(1-2z+2z^2\) \lq\(\frac{1}{1-z}\)_+ \log\frac{\bar{s}}{z\muF^2}
+ 2\(\frac{\log(1-z)}{1-z}\)_+\rq \nonumber \\
&& +\ 2z\ (1-z) \Bigg\}
\lq{\cal B}_{\qb'b}(\bar{s},\bar{Y},\bar{\theta}_1)
+{\cal B}_{qb}(\bar{s},\bar{Y},\bar{\theta}_1)\rq\, ,
\end{eqnarray}
where ${\cal B}_{\qb'b}$ and ${\cal B}_{qb}$ are the corresponding
underlying Born processes.
All the other contributions can be obtained in a similar way.

\subsubsection{Soft and collinear limits of the real contributions}
\label{sec:softcoll_limits}
In the FKS formalism, phase-space singular regions are approached when the
radiation variables $\xi\to 0 $ and/or $y\to \pm 1$. The corresponding
singularities are subtracted from the real cross section using the plus
distributions. One needs to express the singular limits in terms of suitable
radiation variables and of the corresponding underlying Born
contributions. In this section we compute these limits and give explicitly
their expressions.

We start by considering the singular limits of the processes that have
both ISR and FSR singularities, namely $\mathcal{R}_{q q'}$,
$\mathcal{R}_{b q}$ and $\mathcal{R}_{q b}$.  
These processes are the most subtle, being both soft and collinear divergent
for initial- and final-state radiation. As an example, we study
the limits for the $s$-channel scattering $q q' \to t \bar{b}
g$. We can deal with ISR and FSR separately, having defined the
contributions $\mathcal{R}_{qq'}^{3,\splus}$, $\mathcal{R}_{q
q'}^{3,\sminus}$ and $\mathcal{R}_{q q'}^{3,2}$.

For ISR singularities, we use the set $\ISRRad$ to parametrize the
kinematics. When $y\to\pm 1$, the momentum $k_3$ is aligned along the
$\splusminus$ direction and $k_3=\xi\,k_{\splusminus}$, in the CM
frame. The real squared amplitude factorizes and we have
\begin{equation}
\label{eq:coll_lim}
\lq \mathcal{R}_{q q'}^{3,\splusminus}\rq_{y\to\pm 1}=
\frac{4\pi\as}{k_\splusminus\cdot k_3} \,  P^{qq}(z)\, \mathcal{B}_{q q'}
= \CF \frac{1}{\xi^2(1\mp y)} \frac{16\pi\as}{s}\(1+z^2\)
\mathcal{B}_{q q'}\,,
\end{equation}
where $z=(1-\xi)$, $P^{qq}(z)$ is the usual Altarelli-Parisi (AP) splitting
kernel and we have included the real flux factor $1/(2s)$ and a $1/z$ factor
into the $\cal B$ term, as its definition requires. In the FKS approach, one
needs the finite quantity $\xi^2(1\mp
y)\mathcal{R}_{qq'}^{3,\splusminus}$ to perform the subtraction of the
singularities. In the collinear limit, we have
\begin{equation}
\lq\xi^2(1\mp y)\mathcal{R}_{q q'}^{3,\splusminus}\rq_{y=\pm 1}=\CF
\frac{16\pi\as}{s} \(1+z^2\) \mathcal{B}_{q q'}\,.
\end{equation}
In the same limit, we also note that the contributions \mbox{$[\xi^2(1\mp
y)\mathcal{R}_{q q'}^{3,\sminusplus}]$} and  \mbox{$[\xi^2(1\mp
y)\mathcal{R}_{qq'}^{3,2}]$} go to zero, because the factor
$\xi^2(1\mp y)$ makes them finite and the corresponding $\cal{S}$
functions were chosen to vanish in this limit.

In the FSR case, the collinear limit is reached when $y\to 1$. The outgoing
momenta $k_3$ and $k_2$ become parallel and aligned along their sum, denoted
by $k$.  Momentum conservation reads
\begin{equation}
k=k_2 + k_3\,,
\end{equation}
and, in the partonic CM frame, one has
\begin{equation}
k_2=z\,k
\end{equation}
where $z=1-\xi s/(s-\Mrec^2)$. A factorized expression holds in
this case too
\begin{equation}
\lq \mathcal{R}_{q q'}^{3,2}\rq_{y\to 1}=
\frac{4\pi\as}{k_2\cdot k_3}\, P^{qq}(z) \, \mathcal{B}_{q q'}
= \CF \frac{1}{\xi^2(1-y)} \frac{16\pi\as}{zs}\(1+z^2\)
\mathcal{B}_{q q'}\,.
\end{equation}
The finite quantity needed in the application of the subtraction
method is now \mbox{$\xi^2(1-y)\,\mathcal{R}_{qq'}^{3,2}$}, that is
given by
\begin{equation}
\lq\xi^2(1-y)\mathcal{R}_{q q'}^{3,2}\rq_{y=1}=
\CF \frac{16\pi\as}{zs}\(1+z^2\) \mathcal{B}_{q q'}\,.
\end{equation}
We note again that, in this collinear limit, the contributions
$[\xi^2(1-y)\mathcal{R}_{q q'}^{3,\splusminus}]$ vanish,
because of the behaviour of the $\cal{S}$ functions.

The contribution $\mathcal{R}_{qq'}$ is also singular when the outgoing
gluon becomes soft, i.e.~when $k_3\to 0$. In both the two phase-space
parametrizations ($\ISRRad$ and $\FSRRad$), this limit is approached when
$\xi\to 0$.  The Born process has more than 3 colored particles, so that, in
general, one may expect that soft singularities factorize in terms of the
color ordered Born amplitudes~\cite{Frixione:2007vw}.  However, in this case,
the color algebra simplifies, because of the exchange of an intermediate
colorless particle, and we have complete factorization on the Born squared
amplitude. The $\mathcal{R}_{q q'}$ contribution in the soft limit
(eikonal approximation) is given by
\begin{equation}
\lq \mathcal{R}_{q q'}\rq_{\xi\to 0}=8\pi\as \CF
\lg\frac{k_\splus\cdot k_\sminus}{(k_\splus\cdot k_3)(k_\sminus\cdot k_3)} + 
\frac{k_1\cdot k_2}{(k_1\cdot k_3)(k_2\cdot k_3)} -
\frac{M^2}{2(k_1\cdot k_3)^2} \rg \mathcal{B}_{q q'}\,.
\end{equation}
The radiation variable $y$ assumes different meaning in the case of ISR or
FSR (see sec.~\ref{sec:real_kinematics}). In the ISR case, we have the
finite contributions
\begin{equation}
\lq \xi^2 (1\mp y) \mathcal{R}_{qq'}^{3,\splusminus}\rq_{\xi=0} =
4\pi\as \CF
\lg\frac{16}{s(1\pm y)} +
\frac{(s-M^2)(1\mp y)}{(k_1\cdot \hat{k}_3)(k_2\cdot \hat{k}_3)} -
\frac{M^2(1\mp y)}{(k_1\cdot \hat{k}_3)^2} \rg \mathcal{S}^{3,\splusminus}
\,\mathcal{B}_{q q'}\,,
\end{equation}
where $\hat{k}_3=k_3/\xi$ identifies the direction of the soft gluon. In the
FSR case we have instead 
\begin{equation}
\lq \xi^2 (1-y) \mathcal{R}_{qq'}^{3,2}\rq_{\xi=0} =
4\pi\as \CF
\lg\frac{s(1-y)}{(k_\splus\cdot \hat{k}_3)(k_\sminus\cdot \hat{k}_3)} + 
\frac{4(s-M^2)}{(k_1\cdot \hat{k}_3) s \xi_2} -
\frac{M^2 (1-y)}{(k_1\cdot \hat{k}_3)^2} \rg  \mathcal{S}^{3,2}\,
\mathcal{B}_{q q'}\,,
\end{equation}
with $\xi_2=2k_2^0/\sqrt{s}$, defined in the partonic CM frame. 

The $t$-channel processes $\mathcal{R}_{b q}$ and $\mathcal{R}_{q b}$ are
dealt in an analogous way, either for the collinear and the soft limits.
All the other processes have only ISR collinear singularities: the
corresponding limits can be obtained from eq.~(\ref{eq:coll_lim}),
substituting the appropriate AP splitting kernel and the Born term. 

\section{The POWHEG implementation}
\label{sec:powheg_implementation}
\subsection{Generation of the Born variables}
\label{sec:ub_generation}
In the \POWHEG{} method, we first generate the Born kinematics according to
the $\bar{B}$ function, which is the integral of the full NLO cross section
at a given value of the underlying Born kinematics.
It is defined as follows:
\begin{equation}
\bar{B}=\bar{B}_{(s)} + \bar{B}_{(t)}\,,
\end{equation}
where
\begin{equation}
\label{eq:bbarsum_sch}
\bar{B}_{\(s\)}=\sum_{q q'}\bar{B}_{q q'}\,,
\end{equation}
with
\begin{eqnarray}
\label{eq:bbar_sch}
\bar{B}_{q q'}\(\tmmathbf{\bar{\Phi}}_2\)&=&
B_{qq'}\(\tmmathbf{\bar{\Phi}}_2\)  
+ V_{qq'}\(\tmmathbf{\bar{\Phi}}_2\)
+ \int d\FSRRad\ \hat{R}_{q q'}^{3,2}
\(\tmmathbf{\bar{\Phi}}_2,\FSRRad\) \nonumber\\ 
&& + \int d\ISRRad\ \lq \sum_{\splusminus}\hat{R}_{q q'}^{3,\splusminus}
\(\tmmathbf{\bar{\Phi}}_2,\ISRRad\)
+ \hat{R}_{gq, \(s\)} \(\tmmathbf{\bar{\Phi}}_2,\ISRRad\) 
+ \hat{R}_{qg, \(s\)} \(\tmmathbf{\bar{\Phi}}_2,\ISRRad\) \rq \nonumber \\
&& + \int_{\bar{x}_\splus} ^1 \frac{dz}{z}
\lq {G}_{\splus}^{q q'}\({\bf\Phi}_{2,\splus}\) +
    {G}_{\splus}^{gq}\({\bf\Phi}_{2,\splus}\)\rq 
+ \int_{\bar{x}_\sminus} ^1 \frac{dz}{z}
\lq {G}_{\sminus}^{q q'}\({\bf\Phi}_{2,\sminus}\) +
    {G}_{\sminus}^{qg}\({\bf\Phi}_{2,\sminus}\)\rq, \nonumber\\ 
\end{eqnarray}
and where
\begin{equation}
\label{eq:bbarsum_tch}
\bar{B}_{\(t\)}=\sum_{q}\lq \bar{B}_{q b} + \bar{B}_{b q} \rq,
\end{equation}
with
\begin{eqnarray}
\label{eq:bbar_tch1}
\bar{B}_{q b}\(\tmmathbf{\bar{\Phi}}_2\)&=& B_{q b}\(\tmmathbf{\bar{\Phi}}_2\) 
+ V_{q b}\(\tmmathbf{\bar{\Phi}}_2\)
+ \int d\FSRRad\ \hat{R}_{q b}^{3,2} \(\tmmathbf{\bar{\Phi}}_2,\FSRRad\) \nonumber\\
&& + \int d\ISRRad\ \Bigg{[} \sum_{\splusminus} \hat{R}_{q b}^{3,\splusminus}
\(\tmmathbf{\bar{\Phi}}_2,\ISRRad\)
+ \hat{R}_{q g, \(t\)} \(\tmmathbf{\bar{\Phi}}_2,\ISRRad\) \nonumber\\
&& +\ \hat{R}_{g b}^{3,\splus} \(\tmmathbf{\bar{\Phi}}_2,\ISRRad\) 
+ \hat{R}_{g b}^{2,\splus} \(\tmmathbf{\bar{\Phi}}_2,\ISRRad\) \Bigg{]} \nonumber\\
&& + \int_{\bar{x}_\splus} ^1 \frac{dz}{z}
\lq {G}_{\splus}^{q b}\({\bf\Phi}_{2,\splus}\) 
+ {G}_{\splus}^{gb}\({\bf\Phi}_{2,\splus}\)\rq
+ \int_{\bar{x}_\sminus} ^1 \frac{dz}{z}
\lq {G}_{\sminus}^{q b}\({\bf\Phi}_{2,\sminus}\) 
+ {G}_{\sminus}^{qg}\({\bf\Phi}_{2,\sminus}\)\rq\!. \phantom{aaaa}
\end{eqnarray}
The $\bar{B}_{b q}$ contribution can be obtained from
eq.~(\ref{eq:bbar_tch1}) by simply exchanging all flavour indexes and
substituting $\splus \leftrightarrow \sminus$.

According to the \POWHEG{} notation, in eqs.~(\ref{eq:bbar_sch})
and~(\ref{eq:bbar_tch1}) we have traded the $\mathcal{B}$, $\mathcal{V}$,
$\mathcal{R}$ and $\mathcal{G}$ quantities with the corresponding
capital letters, obtained by multiplying them with the appropriate
luminosity $\mathcal{L}$, defined in terms of the parton distribution
functions~(PDF) $f_f^{\splusminus}(\xplusminus, \muF^2)$ as
\begin{equation}
\label{eq:luminosity}
\Lum_{ff'}\!\(\xplus,\xminus\) =  f_f^{\splus}(\xplus, \muF^2)\;
  f_{f'}^{\sminus} (\xminus, \muF^2)\,.
\end{equation}
All the integrals appearing in the above equations are now finite. In
fact, following the FKS subtraction scheme, the hatted functions
\beq
\label{eq:FKS_plusminus}
\hat{\matR}_{ij}^{\splusminus} = \frac{1}{\xi} 
\lg  \(\frac{1}{\xi}\)_{\!\!+} \(\frac{1}{1\mp y}\)_{\!\!+} \rg
\lq \(1\mp y\) \, \xi^2 \, \matR_{ij}^{\splusminus} \rq
\eeq
and
\beq
\label{eq:FKS_out}
\hat{\matR}_{ij}^{\sss\rm FSR} = \frac{1}{\xi} 
\lg  \(\frac{1}{\xi}\)_{\!\!+} \(\frac{1}{1- y}\)_{\!\!+} \rg
\lq \(1- y\) \, \xi^2 \, \matR_{ij}^{\sss\rm FSR} \rq
\eeq
have only integrable divergences when integrated over
$\ISRRad$ and $\FSRRad$ respectively.\footnote{In our case, for both the $s$- and $t$-channels,
\begin{eqnarray}
\hat{\matR}_{ij}^{\splus}&=&\lg \hat{R}_{qq'}^{3,\splus}, 
\hat{R}_{g q, \(s\)},
\hat{R}_{q b}^{3,\splus},
\hat{R}_{g b}^{3,\splus},
\hat{R}_{g b}^{2,\splus},
\hat{R}_{b q}^{3,\splus},
\hat{R}_{g q, \(t\)} \rg\,, \nonumber\\
\hat{\matR}_{ij}^{\sminus}&=&\lg \hat{R}_{qq'}^{3,\sminus}, 
\hat{R}_{q g, \(s\)}, 
\hat{R}_{q b}^{3,\sminus},
\hat{R}_{b g}^{3,\sminus},
\hat{R}_{b g}^{2,\sminus},
\hat{R}_{b q}^{3,\sminus},
\hat{R}_{q g, \(t\)} \rg\,, \nonumber\\
\hat{\matR}_{ij}^{\sss\rm FSR} &=& \lg \hat{R}_{q q'}^{3,2},
\hat{R}_{q b}^{3,2},
\hat{R}_{b q}^{3,2} \rg\,. \nonumber
\end{eqnarray}
}
Some care should still be taken when dealing with the plus distributions.
For more details we refer to refs.~\cite{Frixione:2007vw}
and~\cite{Alioli:2008tz}.

Following ref.~\cite{Frixione:2007vw}, we introduce the $\tilde{B}$
function, defined such that its integral over the radiation variables,
mapped onto a unit cube $\( \lg\xi,y,\phi\rg\to \lg X_{\rm rad}^{(1)},
X_{\rm rad}^{(2)},X_{\rm rad}^{(3)}\rg\)$, gives
\begin{equation}
\bar{B}=\int_0^1\! d^3 X_{\rm rad}\, \tilde{B}\,.
\end{equation}
The generation of the Born variables $\bar{\bf \Phi}_2$ is performed
by using the integrator-unweighter program \MINT{}~\cite{Nason:2007vt}
that, after a single integration of the function $\tilde{B}$ over the
Born and radiation variables, can generate random values for the
variables $\{ \bar{\bf \Phi}_2,X_{\rm rad}\}$, distributed according
to the weight $\tilde{B}\(\bar{\bf\Phi}_2,X_{\rm rad}\)$. We then keep
the $\bar{\bf \Phi}_2$ generated values only, and neglect all the
others, which corresponds to integrate over them.  At this stage, we
also need to choose a Born flavour structure ($\fb$ in the language
of ref.~\cite{Frixione:2007vw}) with a probability proportional to its
relative weight in the $\bar{B}$ function (see
eqs.~(\ref{eq:bbarsum_sch}) and~(\ref{eq:bbarsum_tch})). The event is
then further processed, to generate the radiation variables, as
illustrated in the following section.

\subsection{Generation of the hardest-radiation variables}
\label{sec:rad_generation}
Radiation kinematics is generated using the \POWHEG{} Sudakov form
factor.  For a given underlying Born kinematics
($\tmmathbf{\bar{\Phi}}_2$) and flavour structure ($\fb$), the Sudakov
form factor can be expressed as
\begin{equation}
\Delta^\fb(\tmmathbf{\bar{\Phi}}_2,\pt)=
\prod_{\frindsing\in\{\frindsing|\fb\}} \Delta^\fb_\frindsing
(\tmmathbf{\bar{\Phi}}_2,\pt)\,,   
\end{equation}
where one needs to include in the product all the projected real
contributions that have, as singular limit, the generated underlying Born.
In our case, for the $s$-channel, we can write
\begin{equation}
\label{eq:sudprod_sch}
\Delta^{q q'}(\tmmathbf{\bar{\Phi}}_2,\pt) = 
\Delta^{q q'}_{\sss\rm ISR}(\tmmathbf{\bar{\Phi}}_2,\pt)\
\Delta^{q q'}_{\sss\rm FSR}(\tmmathbf{\bar{\Phi}}_2,\pt)\,,
\end{equation}
where
\begin{eqnarray}
\Delta^{q q'}_{\sss\rm ISR}(\tmmathbf{\bar{\Phi}}_2,\pt)&=&
\exp\Bigg\{
- \int d\ISRRad\,
\frac{\sum_{\splusminus}{R}_{q q'}^{3,\splusminus}
\(\tmmathbf{\Phi_3}\) + {R}_{g q', \(s\)} \(\tmmathbf{\Phi_3}\) 
+ {R}_{qg, \(s\)} \(\tmmathbf{\Phi_3}\)}{B_{q q'}(\tmmathbf{\bar{\Phi}}_2)}  \nonumber\\
&& \qquad\qquad \times\,\theta(\ktISR(\tmmathbf{\Phi_3})-\pt) \Bigg\}
\end{eqnarray}
and
\begin{eqnarray} 
\Delta^{q q'}_{\sss\rm FSR}(\tmmathbf{\bar{\Phi}}_2,\pt)&=&
\exp\lg
- \int d\FSRRad\, 
\frac{{R}_{q q'}^{3,2}
  \(\tmmathbf{\Phi_3}\)}{B_{q q'}(\tmmathbf{\bar{\Phi}}_2)} 
\, \theta(\ktFSR(\tmmathbf{\Phi_3})-\pt)\rg.
\end{eqnarray}
For clarity, here we indicate with ${R}_{g q', \(s\)}$ the real
contribution of $gq$ type that corresponds to the underlying Born $q
q'$. The functions $\ktISR(\tmmathbf{\Phi_3})$ and
$\ktFSR(\tmmathbf{\Phi_3})$ measure the hardness of the radiation in
the real event. In case of ISR singular processes, we chose as
hardness variable the exact transverse momentum of the emitted parton
with respect to the beam axis. In terms of $\ISRRad$, this is given by
\begin{equation}
\label{eq:ktISR}
\ktISR^2=\frac{s}{4}\, \xi^2 \(1-y^2\)=
\frac{\bar{s}}{4(1-\xi)}\, \xi^2 \(1-y^2\).
\end{equation}
For the FSR singular processes, instead, we use as hardness variable
the exact transverse momentum of the FKS parton with respect to the other
light outgoing parton, evaluated in the center-of-mass frame. In terms
of $\FSRRad$, this is given by\footnote{Since for $y\to -1$ no singularities
arise in the FSR case, another possible choice for $\ktFSR$ would be
\begin{equation}
\label{eq:ktFSR2}
\ktFSR^2=\frac{\bar{s}}{2}\, \xi^2 (1-y)\,,
\end{equation}
that has the same behaviour of eq.~(\ref{eq:ktFSR}) in the collinear limit
but has a simpler functional form. We have checked that no sizable
differences arise if one uses eq.~(\ref{eq:ktFSR2}) instead of
eq.~(\ref{eq:ktFSR}).}
\begin{equation}
\label{eq:ktFSR}
\ktFSR^2=\frac{\bar{s}}{4}\xi^2\ (1-y^2)\,.
\end{equation}
The generation of the hardest radiation is performed individually for
$\Delta^{q\bar{q}}_{\sss\rm ISR}$ and $\Delta^{q\bar{q}}_{\sss\rm
FSR}$, and the highest generated $\kt$ is retained.  This corresponds
to generate according to eq.~(\ref{eq:sudprod_sch}), as shown in
Appendix~B of ref.~\cite{Frixione:2007vw}.  If $\kt$ is below a given
cut, $\ptmin$, no radiation is generated, and a Born event is
returned.

The upper bounding functions for
the application of the veto method have been chosen in the following way:
\begin{equation}
\label{eq:uboundISR}
\frac{\sum_{\splusminus}{R}_{q q'}^{3,\splusminus}
\(\tmmathbf{\Phi_3}\) + {R}_{g q', \(s\)} \(\tmmathbf{\Phi_3}\) 
+ {R}_{qg, \(s\)} \(\tmmathbf{\Phi_3}\)}{B_{q q'}(\tmmathbf{\bar{\Phi}}_2)}
\, J_{\rm rad}^{\sss\rm ISR}(\tmmathbf{\bar{\Phi}}_2,\ISRRad)
\le N_{q q'}^{\sss\rm ISR} \, \frac{\as(\ktISR^2)}{\xi\,(1-y^2)}
\end{equation}
for ISR, and
\begin{equation}
\label{eq:uboundFSR}
\frac{{R}_{q q'}^{3,2} \(\tmmathbf{\Phi_3}\)}
{B_{q q'}(\tmmathbf{\bar{\Phi}}_2)}\, J_{\rm rad}^{\sss\rm
  FSR}(\tmmathbf{\bar{\Phi}}_2,\FSRRad) 
\le N_{q q'}^{\sss\rm FSR} \, \frac{\as(\ktFSR^2)}{\xi\,(1-y^2)}
\end{equation}
for FSR.  

The same procedures holds also for the $t$-channel case, with
appropriate modifications in
formulae~(\ref{eq:sudprod_sch})--(\ref{eq:uboundFSR}).

The method used to generate radiation events according to these
upper bounding functions is analogous to the one described in
Appendix~D of ref.~\cite{Nason:2006hfa}, and we do not repeat it
here.

As a final remark, we also point out that single-top $s$- and $t$-channel
Born cross sections vanish at some points in the Born phase space, as one can
argue by looking at eqs.~(\ref{eq:b_ud})--(\ref{eq:b_bd_db}). For this
reason, special care has to be taken during the radiation generation
procedure. We handled this problem using the same method described in
sec.~3.3 of ref.~\cite{Alioli:2008gx}. We thus refer to that paper for
further details.

\subsection{Top-quark decay}
\label{sec:top_decay}
The calculation we have described so far leads to the generation of events
with an undecayed top quark. We include the decay kinematics effects
in an approximate way, by requiring that the decay products are distributed
with a probability proportional to the tree-level cross section for the full
production and decay process. This procedure was first suggested in
ref.~\cite{Frixione:2007zp}.  In the following we describe our
implementation, focusing upon the decay $t \to b W^+ \to b\bar{\ell}\nu $.

We first generate a Born-like or real-like event according to the \POWHEG{}
method.  In both cases we denote the set of
variables that parametrize the undecayed momenta as $\powPS$ and the
corresponding flavour structure as $f$.  As described at the end of
sec.~\ref{sec:Born_kinematics}, at this stage the top virtuality $M^2$ is
distributed according to a Breit-Wigner function.  We write the tree-level
cross section for production and decay in the following form 
\begin{equation}
\label{eq:full_amp2}
d\sigma^f_{\rm dec}=\frac{1}{2s}\, 
\Lum\, \mathcal{M}^f_{\rm dec}(\powPS,\decPS) \, d\Phi_{\rm dec}\,,
\end{equation}
where $\Lum$ is the luminosity factor and $\mathcal{M}^f_{\rm dec}$ is
the squared amplitude corresponding to the full decayed process that
originates from the undecayed process $f$.\footnote{The full
tree-level squared amplitudes $\mathcal{M}^f_{\rm dec}$ have been
obtained using MadGraph.} For consistency,
the squared amplitude $\mathcal{M}^f_{\rm dec}$ must include only
resonant graphs (i.e.~graphs where the top momentum equals the sum of the
$b$, $\bar{\ell}$ and $\nu$ momenta).  
We write the full phase
space, including the decay, in the factorized form
\begin{equation}
d\Phi_{\rm dec}= d\powPS \, d\decPS\,,
\end{equation}
where $\powPS$ is the undecayed (\POWHEG{}) phase space and $\decPS$
is defined implicitly by this equation. We notice that
\begin{equation}
\mathcal{M}^f_{\rm undec} \times \mbox{BR}(t\to b \bar{\ell} \nu) = 
\int \mathcal{M}^f_{\rm dec}\ d\decPS\,,
\end{equation}
where $\mathcal{M}^f_{\rm undec}$ is the undecayed squared amplitude,
i.e.~the Born or real amplitude that we used throughout the
computation.
Thus, the differential probability
$dP(\decPS | \powPS)$ for the generation of $\decPS$ from a given undecayed
kinematics $\powPS$ is 
\begin{equation}
\label{eq:dec_prob}
dP(\decPS | \powPS) =\frac{1}{\mbox{BR}(t\to b \bar{\ell} \nu)}
\, \frac{\mathcal{M}^f_{\rm dec}(\powPS,\decPS)} 
{\mathcal{M}^f_{\rm undec}(\powPS)} \, d\decPS\,.
\end{equation}
To generate efficiently $\decPS$ distributed according to~(\ref{eq:dec_prob})
we use the hit-and-miss technique and so we need to find an upper bounding
function for $dP$. This bound can be guessed from the structure of the top
decay. In our case, we use as upper
bound for the ratio $\mathcal{M}^f_{\rm
  dec}(\powPS,\decPS)/\mathcal{M}^f_{\rm undec}(\powPS)$, the expression
\begin{equation}
\label{eq:ub_dec}
U_{\rm dec} (M^2,\decPS)= N_{\rm dec} \
\frac{\mathcal{M}_{t\to b W}(M^2,M_{\bar{\ell} \nu}^2)}
{(M^2-m_t^2)^2 + m_t^2 \, \Gamma_t^2}
\ \frac{\mathcal{M}_{W\to \bar{\ell} \nu}(M_{\bar{\ell} \nu}^2)}
{(M_{\bar{\ell} \nu}^2-m_W^2)^2 + m_W^2 \, \Gamma_W^2}\,,
\end{equation}
where $M_{\bar{\ell} \nu}^2=(k_{\bar{\ell}}+k_{\nu})^2$ and
$\mathcal{M}_{t\to b W}$ and $\mathcal{M}_{W\to \bar{\ell} \nu}$ are the
decay squared amplitudes corresponding to the subprocesses in their
subscripts.  In the previous formula, as well as in $\mathcal{M}^f_{\rm
dec}$, finite-width effects have been fully taken into account. One can
predict the appropriate value for the normalization factor $N_{\rm dec}$ as
explained in ref.~\cite{Frixione:2007zp} or compute it by sampling the decay
phase space $\decPS$ and comparing $U_{\rm dec}$ with the exact expression,
in such a way that the inequality
\begin{equation}
\mathcal{M}^f_{\rm dec}(\powPS,\decPS) \le
\mathcal{M}^f_{\rm undec}(\powPS)\ U_{\rm dec}(M^2,\decPS)
\end{equation}
holds. The veto algorithm is then applied:
\begin{enumerate}
\item 
\label{item:veto}
First one generates a point in the phase space  $\decPS$.
\item Then a random number $r$ in the range $\lq 0,
U_{\rm dec} (M^2,\decPS) \rq$ is generated.
\item If $r< \mathcal{M}^f_{\rm dec}(\powPS,\decPS) / \mathcal{M}^f_{\rm
undec}(\powPS)$, keep the decay kinematics and generate the event. Otherwise
go back to step~\ref{item:veto}.
\end{enumerate}

\section{Results}
\label{sec:results}
In this section we present our results and comparisons with the fixed
order (next-to-leading) calculation and with the \MCatNLO{}~3.3 and
\PYTHIA{}~6.4.21 Shower Monte Carlo~(SMC) programs.\footnote{This
newest update of \PYTHIA{} yields more consistent results when
multiple interactions are turned on in user-initiated processes
(see the release notes in \url{http://projects.hepforge.org/pythia6/}).}
We have used the CTEQ6M~\cite{Pumplin:2002vw} set
for the parton distribution functions and the associated value of
\mbox{$\Lambda_{\scriptscriptstyle\overline{\rm
MS}}^{(5)}=0.226$~GeV}.  Furthermore, as discussed in
refs.~\cite{Frixione:2007vw,Nason:2006hfa}, we use a rescaled value
\mbox{$\Lambda_{\scriptscriptstyle\rm{MC}}=
1.569\,\Lambda_{\scriptscriptstyle\overline{\rm MS}}^{(5)}$} in the
expression for $\as$ appearing in the Sudakov form factors, in order
to achieve next-to-leading logarithmic accuracy.

Although the matrix-element calculation has been performed in the
massless-quark limit (except, of course, for the top quark), the lower cutoff
in the generation of the radiation has been fixed according to the mass of
the emitting quark. The lower bound on the transverse momentum for the
emission off a massless emitter ($u$, $d$, $s$) has been set to the value
$\ptmin = \sqrt{5}\,\Lambda_{\scriptscriptstyle\rm{MC}}$. We instead choose
$\ptmin$ equal to $m_c$ or $m_b$ when the gluon is emitted by a charm or a
bottom quark, respectively.  We set $m_c=1.55$~GeV and $m_b= 4.95$~GeV.

The renormalization and factorization scales have been taken equal to the
radiated transverse momentum during the generation of radiation (see
eqs.~(\ref{eq:ktISR}) and~(\ref{eq:ktFSR})), as the \POWHEG{} method
requires. We have also taken into account properly the heavy-flavour
thresholds in the running of $\as$ and in the PDF's, by changing the number
of active flavours when the renormalization or factorization scales cross a
mass threshold.  In the $\bar{B}$ calculation, instead, $\mu_R$ and $\mu_F$
have been chosen equal to the top-quark mass, whose value has been fixed to
$m_t=175$~GeV.  In all the comparisons, we have kept the top-quark
virtuality $M^2$ fixed to $m_t^2$, so that matrix elements have been
evaluated assuming $\Gamma_t=0$. We have also set $\Gamma_W=0$ in all the
propagators.  The other relevant parameters are
\begin{equation}
M_W= 80.4 \mbox{ GeV}\,,\qquad \sin^2\theta_W^{\rm eff}=0.23113\,,
\qquad \alpha^{-1}_{\rm em}(m_t)=127.011989\,.
\end{equation}
From the above values, the weak coupling has been computed as
$g=\sqrt{4\pi\alpha_{\rm em}}/\sin\theta_W^{\rm eff}$.
In addition, for sake of comparison, we fixed the CKM matrix elements equal to
\begin{equation}
\begin{array}{c}
\\
V_{\rm \sss CKM}=
\end{array}
\begin{array}{c c}
& d\quad\quad\ \ s\ \ \quad\quad b \\
\begin{array}{c}
u\\
c\\
t
\end{array} 
&
\left(
\begin{array}{c c c}
0.9740 & 0.2225 & 0.0000\\
0.2225 & 0.9740 & 0.0000\\
0.0000 & 0.0000 & 1.0000
\end{array}
\right).
\end{array}
\end{equation}
In order to minimize effects due to differences in the shower and
hadronization algorithms, we have interfaced \POWHEG{} with the
\HERWIG{} angular-ordered shower when comparing with \MCatNLO{} and
with the $\pt$-ordered \PYTHIA{} shower when comparisons with
\PYTHIA{} have been carried out.

All the following results have been obtained assuming that the top decays
semileptonically ($t\to b\,\bar{\ell}\,\nu$), as explained in
sec.~\ref{sec:top_decay}, but removing the branching ratio, so that plots are
normalized to the total cross section.

We present a few distributions, done mainly for comparison with \MCatNLO{}
and with the NLO calculation.  Some of them are ``unphysical'', i.e., for
example, when talking of the top-quark momentum $p^t$, we refer to the exact
$p^t$ taken directly from the MC shower history, right before the top decay.
For sake of simplicity, we also force the lightest $b$-flavoured hadrons to be
stable after the hadronization stage of SMC programs.

Jets have been defined according to the $\kt$
algorithm~\cite{Catani:1993hr}, as implemented in the {\tt FASTJET}
package~\cite{Cacciari:2005hq}, setting $R=1$ and imposing a lower
$10$~GeV cut on jet transverse momenta.  We call ``top jet'' the jet
that contains the hardest $b$-flavoured hadron,\footnote{Here we mean
precisely $b$-flavoured, i.e. not $\bar{b}$-flavoured, that arises in
the production process.} which will, most of the time, come from the
top-quark decay.  The other reconstructed jets will come from the
shower of massless partons, and we call them ``light
jets''.\footnote{In the fixed-order calculation, instead, the top
quark is not decayed, and the top jet corresponds to the jet that
contains the top quark.}  In this way, the momentum $p^t$ of the top
quark and the momentum of the top jet are different, since the last
may or may not include all the particles from the top decay and
shower.

\subsection{Tevatron results}  
\begin{figure}[!htb]
\begin{center}
\epsfig{file=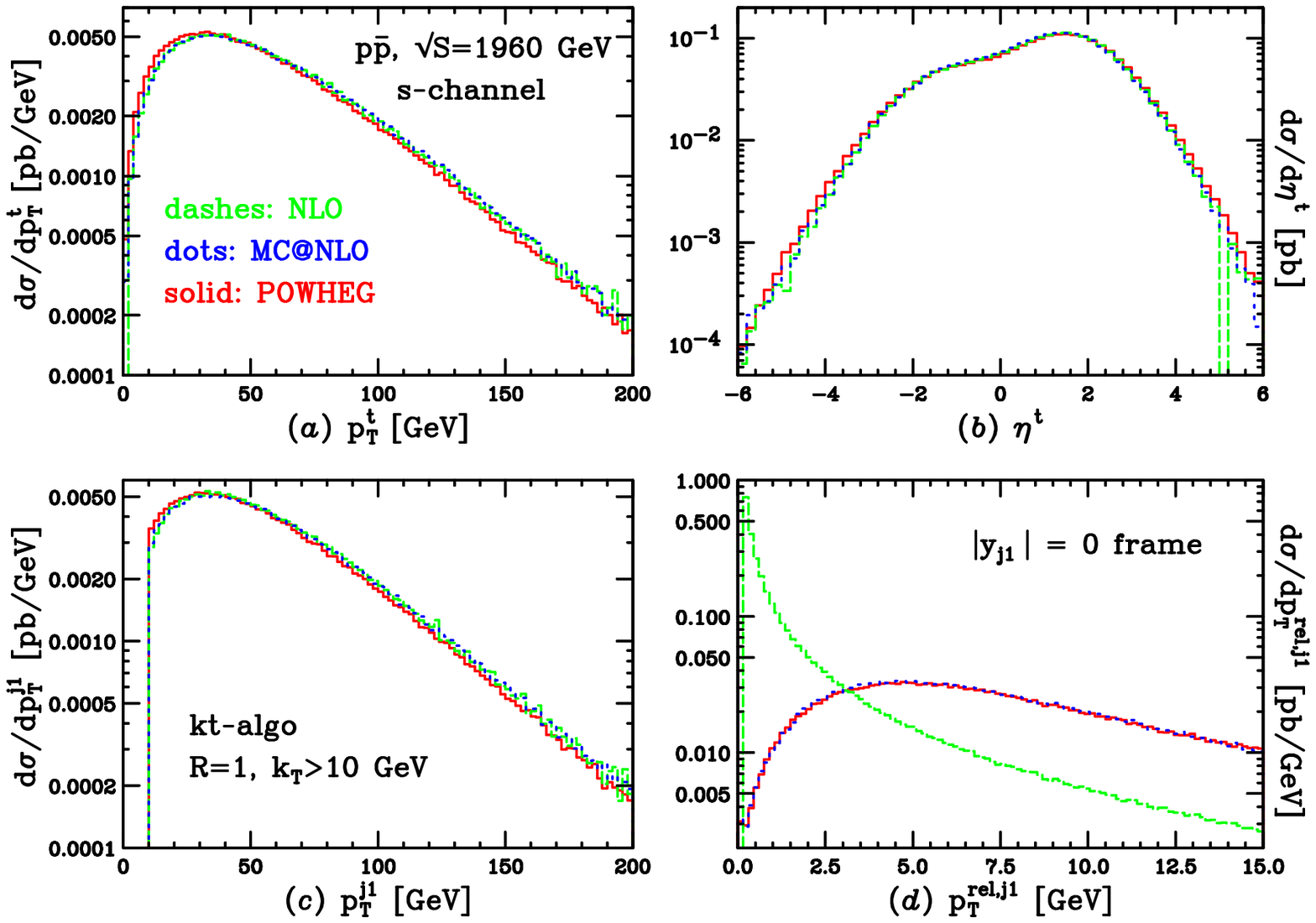,width=\figwidth}\\~\\
\epsfig{file=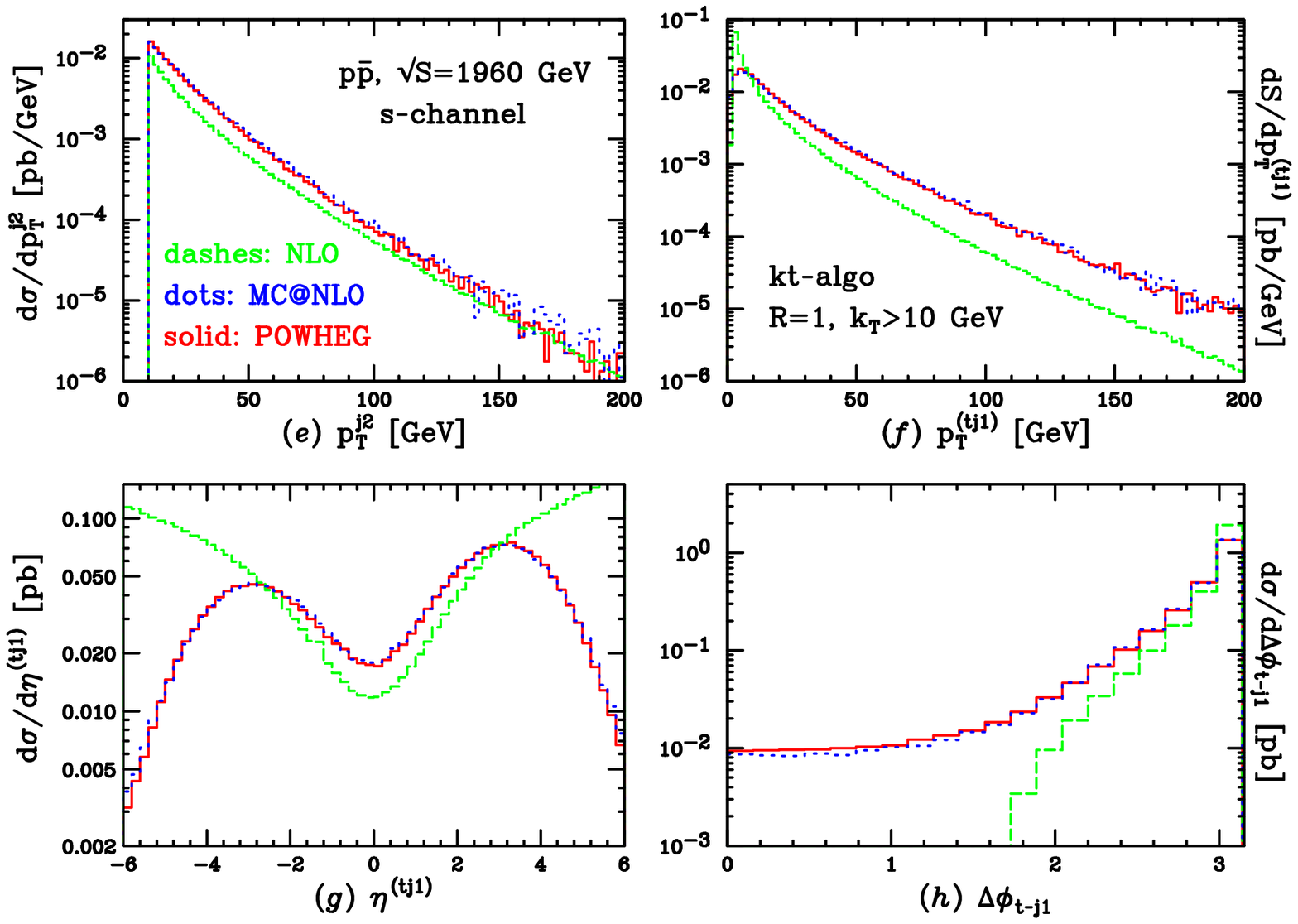,width=\figwidth}
\end{center}
\captskip
\caption{\label{fig:cmp_s_tev-mcatnlo} Comparisons between \POWHEG{},
\MCatNLO{} and NLO results for $s$-channel top production at the
Tevatron $p\bar{p}$ collider.}
\end{figure}

We start comparing various kinematical variables for single-top $s$-channel
production at the Tevatron $p\bar{p}$ collider.  In
fig.~\ref{fig:cmp_s_tev-mcatnlo} we have collected the following
distributions:
\begin{itemize}
\item In panels~(\emph{a}) and~(\emph{b}) we show the transverse
  momentum $\pt^{\,t}$ and the pseudorapidity $\eta^{\,t}$ of the top quark and
   in panel~(\emph{c}) we show the hardest jet transverse momentum
  $\pt^{\,j_1}$.  The agreement with the
  fixed-order calculation and with the \MCatNLO{} results is very good. Only
  the top transverse-momentum distribution shows a tiny mismatch, our result
  being slightly softer than the NLO and the \MCatNLO{} ones. When
  interfacing \POWHEG{} with \PYTHIA{}, we instead find full overlapping with
  the NLO result. It is thus likely that this small feature may be attributed
  to shower effects.
\item In panel~(\emph{d}), we plot $\pt^{{\rm rel},j_1}$, the relative
  transverse momentum of all the particles clustered inside the hardest
  jet. This is defined as follows:
\begin{itemize}
\item We perform a longitudinal boost to a frame where the hardest-jet
  rapidity is zero.
\item In this frame, we compute the quantity
\begin{equation}
\pt^{{\rm rel},j_1}=\sum_{i\in j_1} \frac{| \vec{k}^i \times
  \vec{p}^{\, j_1}|}{|\vec{p}^{\, j_1}|}\,,
\end{equation}
where $k^i$'s are the momenta of the particles that belong to the hardest jet
that, in this frame, has momentum $p^{\,j_1}$.
\end{itemize}

This quantity is thus the sum of the absolute values of the transverse
momenta, taken with respect to the jet axis, of the particles inside the
hardest jet, in the frame specified above.  The plot shows a marked
disagreement between fixed order calculation and showered results. This
disagreement is well understood, since the observable we are considering is a
measure of the spreading of the hardest jet. Thus, its shape is strongly
affected by the Sudakov form factor and it is well described by SMC
programs. The NLO calculation cannot give, instead, a reliable estimate, since
when $\pt^{{\rm rel},j_1}\to 0$ the differential cross section diverges.

\item In plots~(\emph{e}) and~(\emph{f}), the next-to-hardest jet transverse
momentum $\pt^{\,j_2}$, and the transverse momentum of the system made by the
top quark and the hardest jet, $\pt^{(t j_1\!)}$, are shown.  We see a
remarkable good agreement between our program and \MCatNLO{}, while sensible
differences with respect to the NLO results are present.  At the NLO parton
level, $\pt^{\,j_2}$ and $\pt^{(t j_1\!)}$ balance against each other, so
that the two distributions coincide down to the minimum $\pt$ cut present in
the first plot.

In plot~(\emph{e}), we see an enhancement of the showered results at
intermediate values of $\pt$, while in plot~(\emph{f}) we see a low-$\pt$
suppression and an enhancement at intermediate and high $\pt$.  The low-$\pt$
suppression is clearly a Sudakov effect.  The high-$\pt$ enhancement comes
instead from events in which the hardest parton is well balanced against the
top quark, but where many hadrons, coming from the hardest parton, end up in
the top jet, and are thus removed, or they end up out of the jet
cluster. This creates an artificial imbalance, and thus an effective $\pt$
for the $(t j_1\!)$ system. These effects are so pronounced because the cross
section for a balanced top-quark--hardest-jet system is much higher, since it
does not require the production of an additional hard parton.  We have
verified this hypothesis by analyzing \POWHEG{} outputs before the showering
stage, either clustering or not the $b$ quark coming from the top decay. In
the case where the $b$ quark is included in the analysis (and the jet
containing the $b$ is removed from the jet sample), we see a marked rise of
the $\pt$ tail. A further rise is observed when the shower is turned on, and
may be attributed to energy lost out of the hardest jet cluster due to
showering.  We see no such effect for the next-to-hardest jet spectrum in
plot~(\emph{e}). There, the raise at medium $\pt$ may be attributed to the
shower $\pt$ smearing.

\item Finally, in plots~(\emph{g}) and~(\emph{h}), the pseudorapidity
$\eta^{(t j_1\!)}$ of the top-quark--hardest-jet system and the azimuthal
difference $\Delta\phi_{t\mbox{-}j_1}=|\phi_t-\phi_{j_1}|$ are shown.  The
pseudorapidity of the $(t j_1\!)$ system shows an expected discrepancy
between the showered results and the fixed order one: radiation near the beam
axis is suppressed by the Sudakov form factor but not in the NLO result,
giving rise to the higher tails at large $|\eta^{(t j_1\!)}|$. In
plot~(\emph{h}), \MCatNLO{} and \POWHEG{} differ instead from the fixed order
result for a kinematical reason: at the parton level, having at most three
particles, there is no phase space for the next-to-hardest jet to recoil
against the $(t j_1\!)$ system when $\Delta\phi_{t\mbox{-}j_1}<\pi/2$.
\end{itemize}

\begin{figure}[htb]
\begin{center}
\epsfig{file=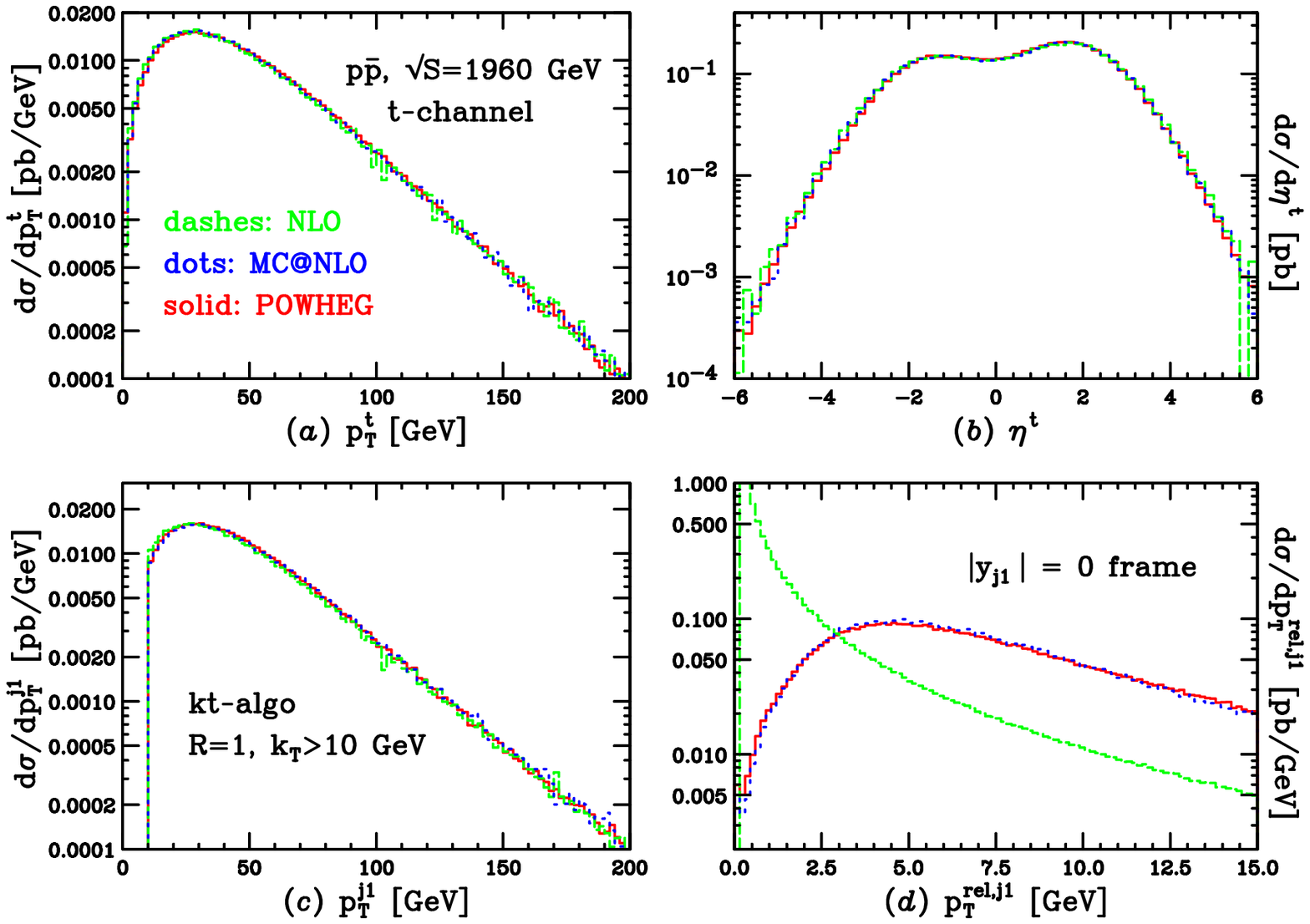,width=\figwidth}\\~\\
\epsfig{file=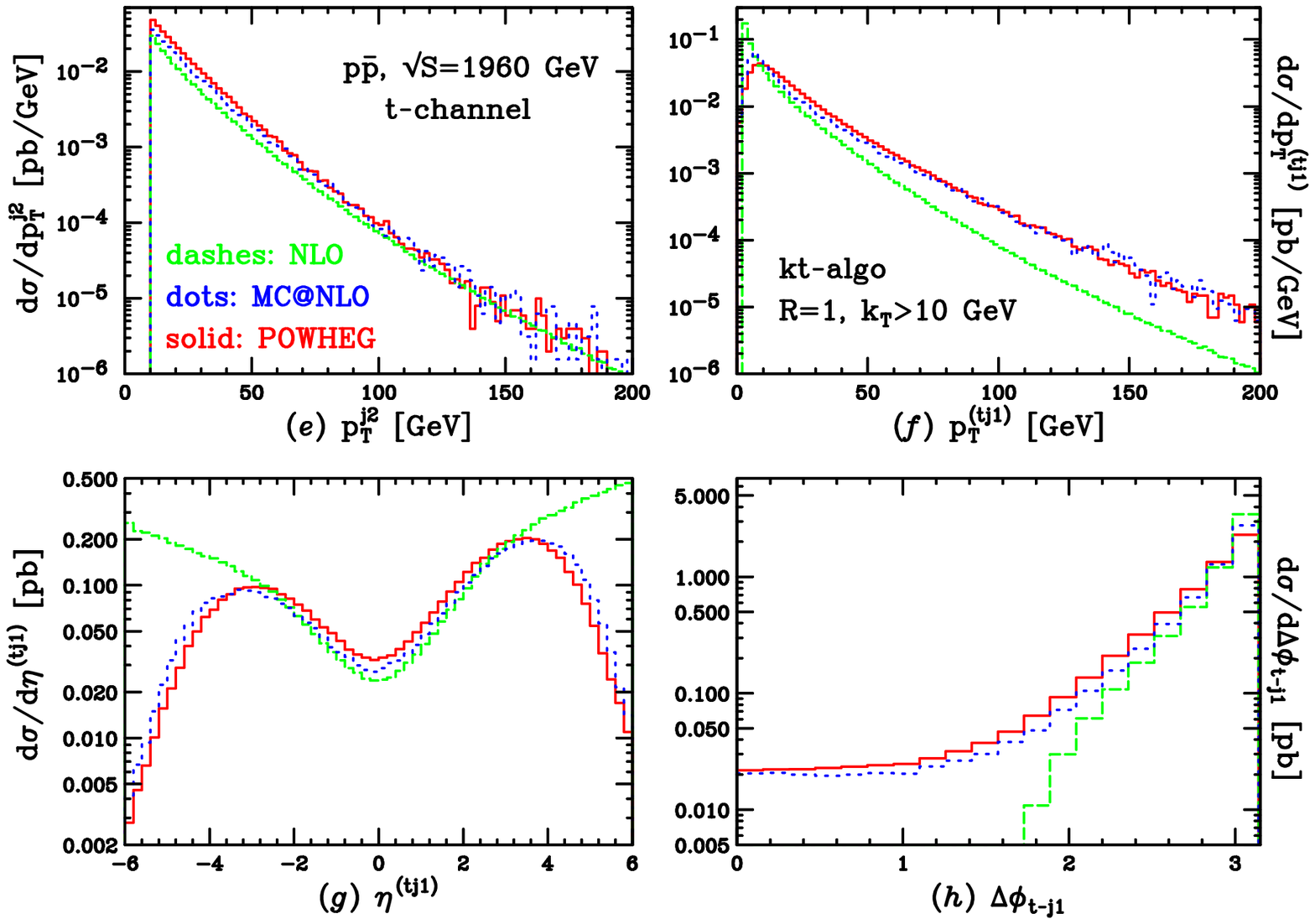,width=\figwidth}
\end{center}
\captskip
\caption{\label{fig:cmp_t_tev-mcatnlo}
Comparisons between \POWHEG{}, \MCatNLO{} and NLO results for
$t$-channel top production at the Tevatron $p\bar{p}$ collider.}
\end{figure}

\begin{figure}[htb]
\begin{center}
\epsfig{file=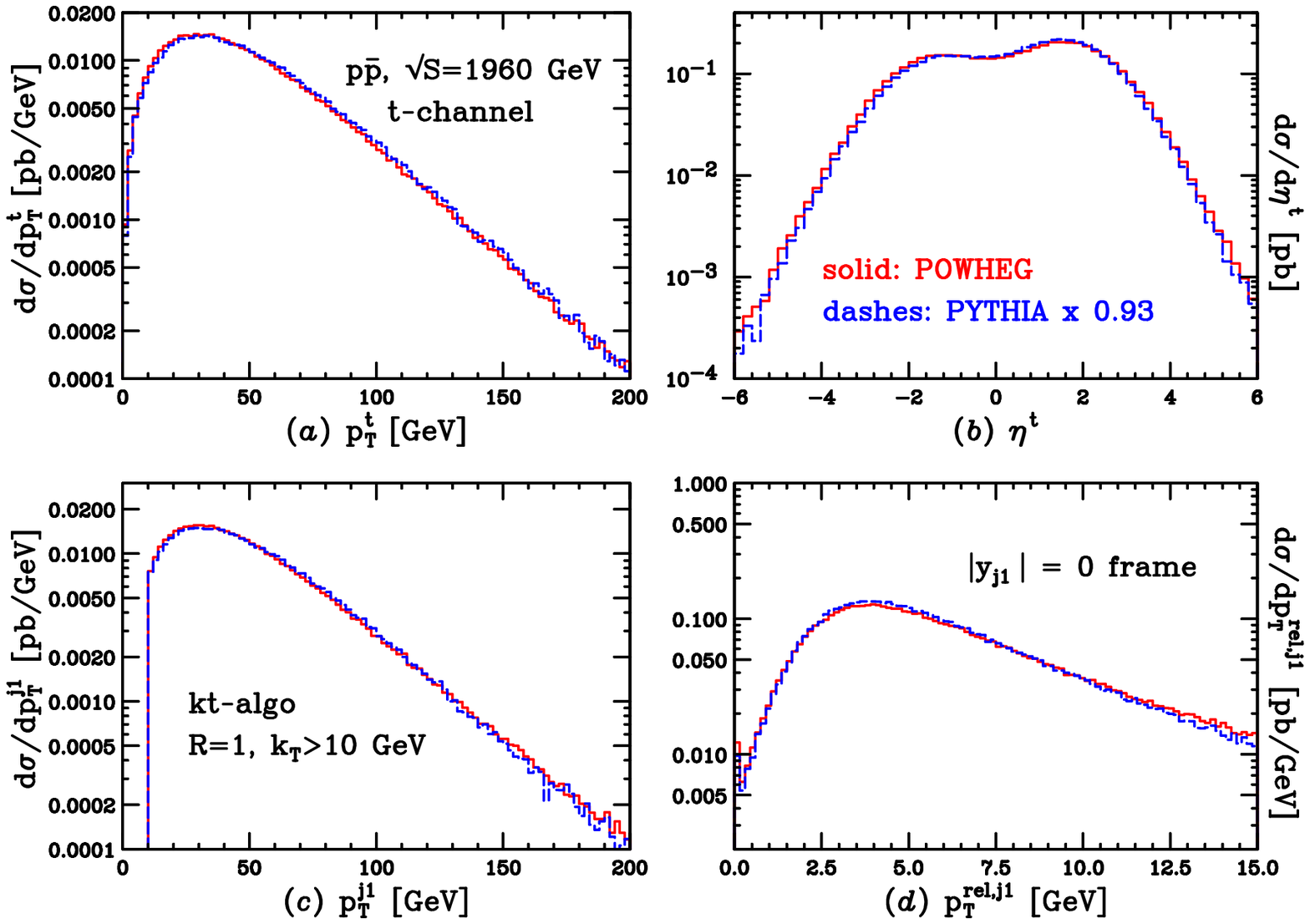,width=\figwidth}\\~\\
\epsfig{file=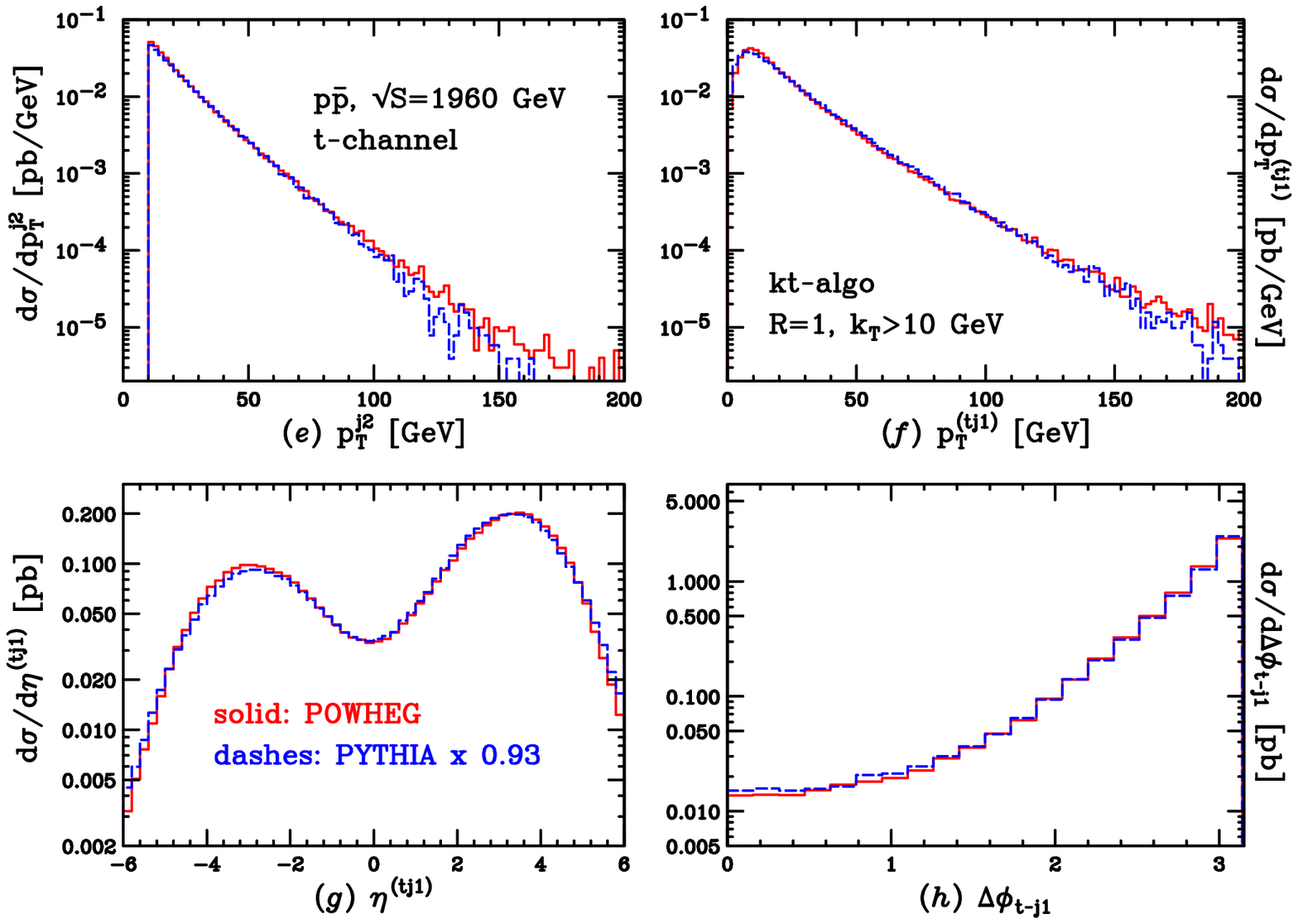,width=\figwidth}
\end{center}
\captskip
\caption{\label{fig:cmp_t_tev-pythia}
Comparisons between \POWHEG{} and \PYTHIA{} results for
$t$-channel top production at the Tevatron $p\bar{p}$ collider.}
\end{figure}
\cleardoublepage

A similar set of comparisons is presented in fig.~\ref{fig:cmp_t_tev-mcatnlo}
for the $t$-channel production mechanism, always at the Tevatron. The
agreement between \POWHEG{} and \MCatNLO{} is as good as before for inclusive
quantities, or even better. In particular, the slight mismatch in the top
transverse-momentum distribution completely disappears, as one can see in
plot~(\emph{a}). For all the other plots,
considerations similar to the $s$-channel case remain valid.

In fig.~\ref{fig:cmp_t_tev-pythia} the same set of plots are shown,
comparing \POWHEG{} and \PYTHIA{}. We have good agreement for most
distributions, after applying an appropriate $K$ factor to the
\PYTHIA{} results. Only minor differences are present in the
high-$\pt$ tail of distributions in panels~(\emph{e}) and~(\emph{f}).

\begin{figure}[htb]
\begin{center}
\epsfig{file=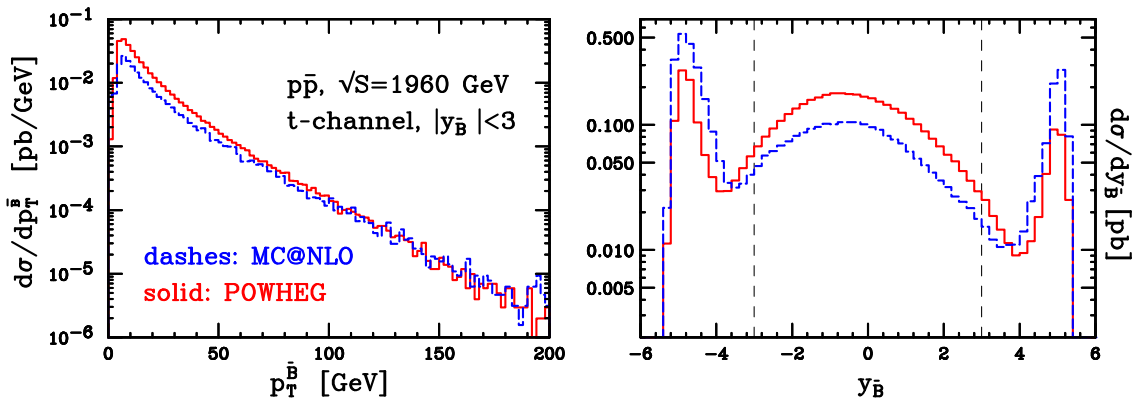,width=\figwidth}
\end{center}
\captskip
\caption{\label{fig:bbar_t_tev-mcnlo} Comparisons between \POWHEG{}
and \MCatNLO{} results for the hardest $\bar{b}$-flavoured
hadron transverse momentum (left) and rapidity (right), for
$t$-channel top production at the Tevatron $p\bar{p}$
collider. Rapidity cuts are highlighted.}
\end{figure}

As a final comparison, in the left panel of fig.~\ref{fig:bbar_t_tev-mcnlo},
we show $\pt^{\bar{B}}$, the transverse-momentum spectrum of the hardest
$\bar{b}$-flavoured hadron, after imposing the rapidity cut
$|y_{\bar{B}}|<3$.  In the $t$-channel, this hadron will come most probably
from an initial-state gluon undergoing a $b\bar{b}$ splitting. The $b$ quark
is then turned into a $t$ while the $\bar{b}$ quark is showered and
hadronized.  We see that, while \POWHEG{} and \MCatNLO{} are in a fair
agreement in the medium- and high-$\pt$ range, sizable differences are
present at low $\pt$.  These discrepancies are most probably due to the
disagreement that one can notice in the $y_{\bar{B}}$ distribution (right
panel of fig.~\ref{fig:bbar_t_tev-mcnlo}), and to a smaller extent to a
different implementation of the inclusion of $b$-mass effects by both
programs (just before the showering stage).

\begin{figure}[htb]
\begin{center}
\epsfig{file=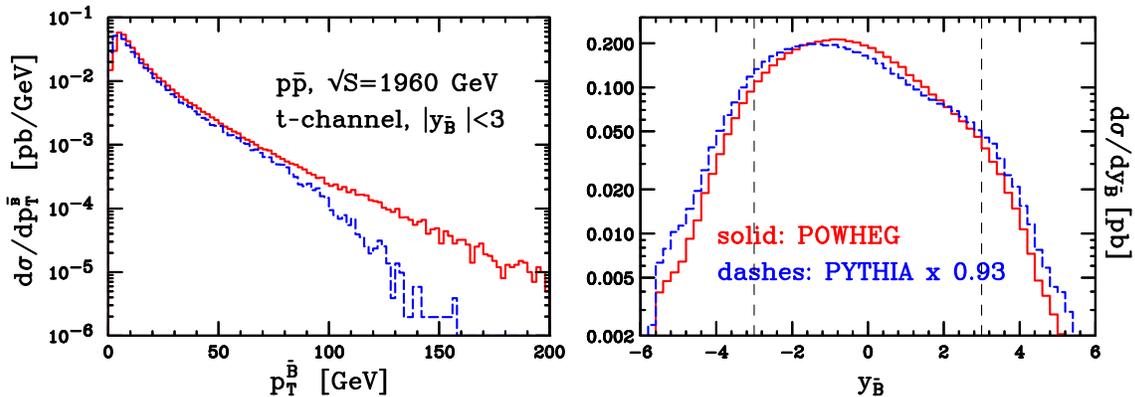,width=\figwidth}
\end{center}
\captskip
\caption{\label{fig:bbar_t_tev-pythia} Comparisons between \POWHEG{}
and \PYTHIA{} results for the hardest $\bar{b}$-flavoured
hadron transverse momentum (left) and rapidity (right), for
$t$-channel top production at the Tevatron $p\bar{p}$
collider. Rapidity cuts are highlighted.}
\end{figure}

We also plot in fig.~\ref{fig:bbar_t_tev-pythia} the same quantities
comparing \POWHEG{} interfaced to \PYTHIA{} with respect to \PYTHIA{}
alone.  A large mismatch in the high-$\pt^{\bar{B}}$ spectrum is
clearly visible in the left panel. This observable is particularly
sensitive to real matrix-element effects, not present in \PYTHIA{}.
Concerning the low-$\pt^{\bar{B}}$ behaviour, we see that here the
difference is much less pronounced than in fig.~\ref{fig:bbar_t_tev-mcnlo}.
Furthermore, the aforementioned mismatch in the $y_{\bar{B}}$
distribution is no longer present, as one can see in the right panel.

By comparing figs.~\ref{fig:bbar_t_tev-mcnlo}
and~\ref{fig:bbar_t_tev-pythia}, one immediately notices the different
behaviours of the two Monte Carlo programs that we are interfacing to.  We
observe that the \HERWIG{} shower and hadronization create an enhancement at
large values of $|y_{\bar{B}}|$, which is not present in \PYTHIA{}.  This
feature is known to the \HERWIG{} authors,\footnote{See M.~Seymour's talk in
\url{http://bwhcphysics.lbl.gov/vplusjets.html}.} and is traced back to a
mismatch of the scale at which backward evolution is switched off, with the
scale at which the $b$-quark density is turned on in the pdf's.  The effect
is more pronounced in \MCatNLO{}, probably due to the fact that \POWHEG{}
does not rely on \HERWIG{} for the generation of the hardest splitting.

\subsection{LHC results}
\begin{figure}[htb]
\begin{center}
\epsfig{file=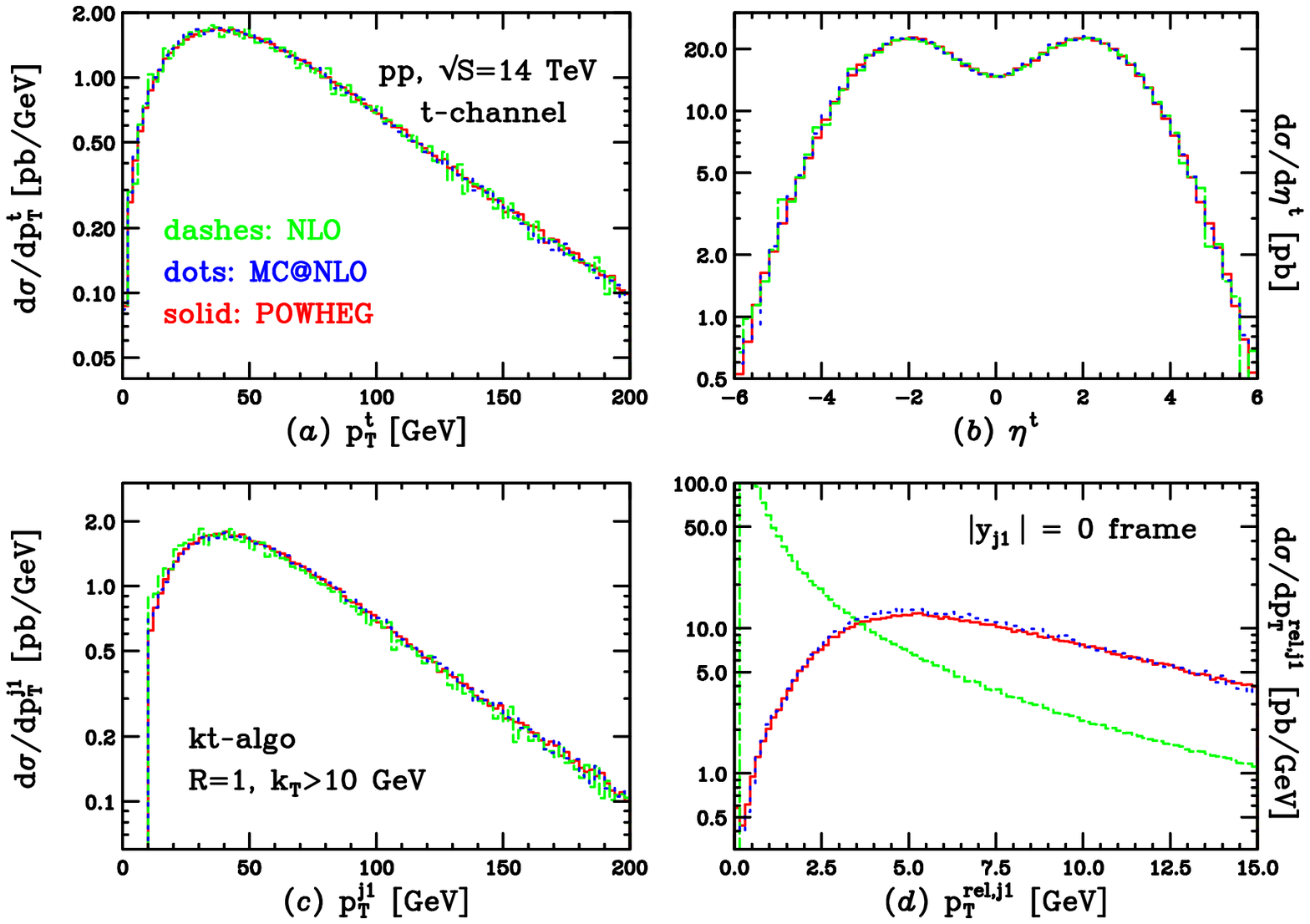,width=\figwidth}\\~\\
\epsfig{file=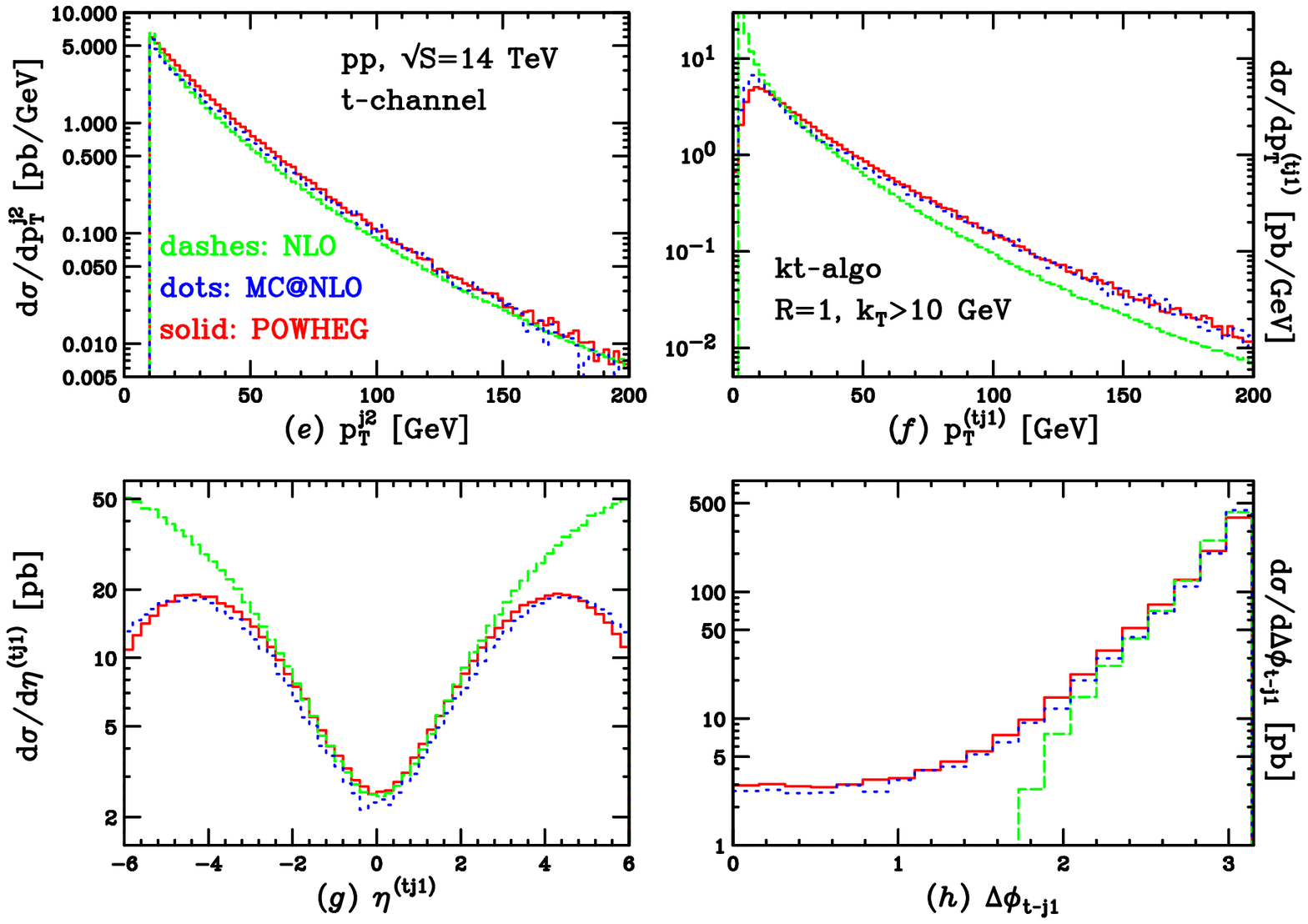,width=\figwidth}
\end{center}
\captskip
\caption{\label{fig:cmp_t_LHC-mcatnlo}
Comparisons between \POWHEG{}, \MCatNLO{} and NLO results for
$t$-channel top production at the LHC $pp$ collider.}
\end{figure}

\begin{figure}[htb]
\begin{center}
\epsfig{file=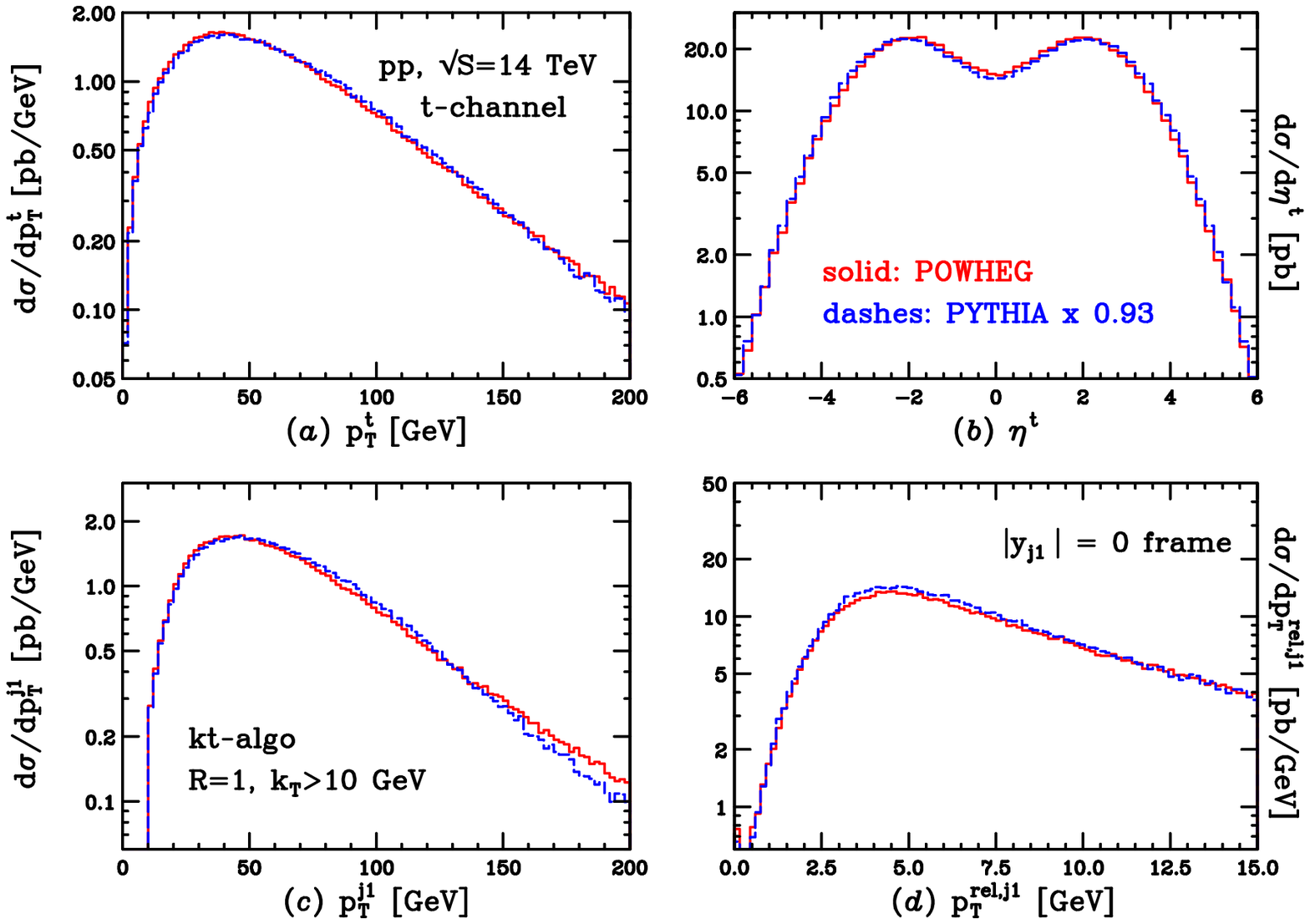,width=\figwidth}\\~\\
\epsfig{file=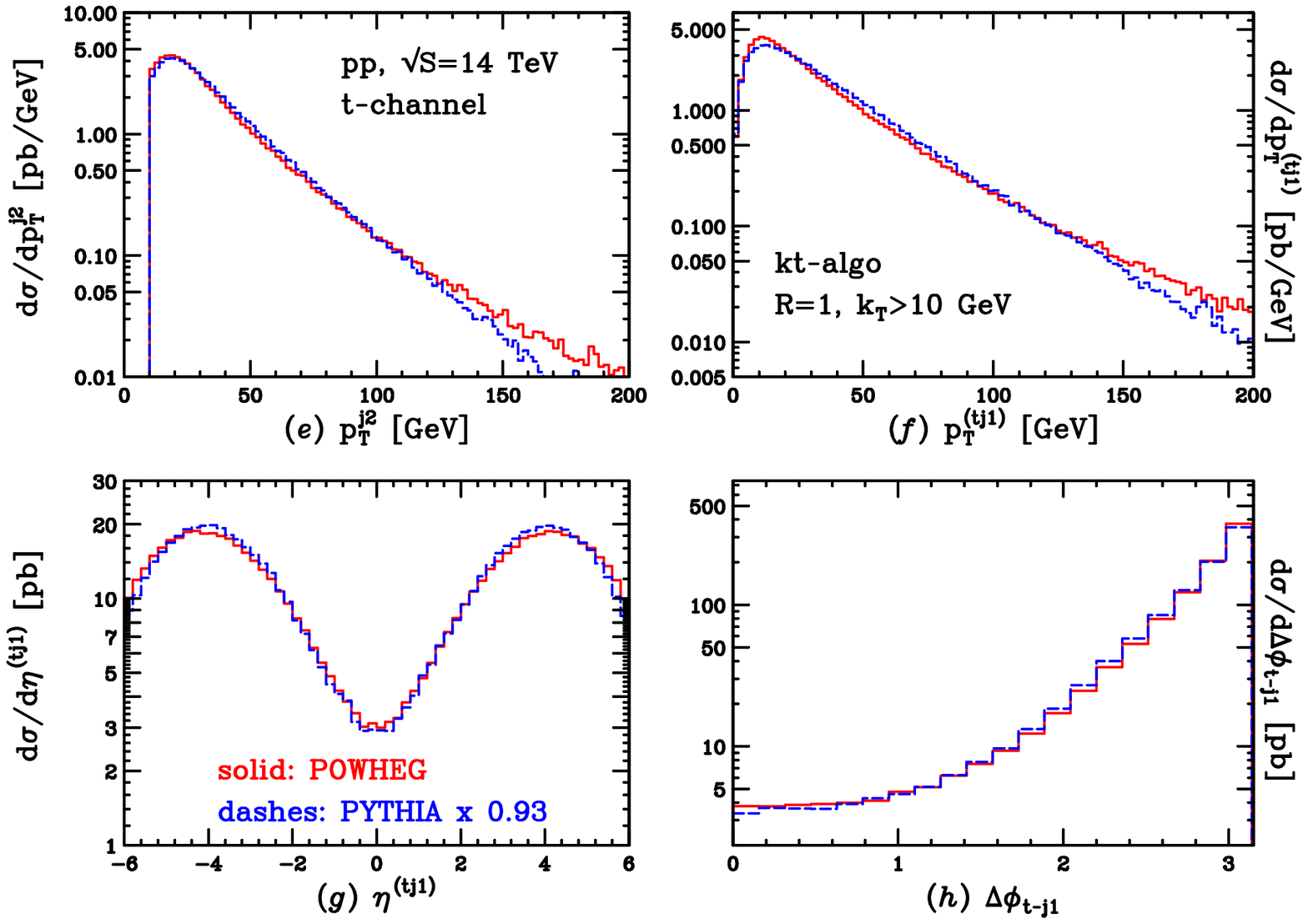,width=\figwidth}
\end{center}
\captskip
\caption{\label{fig:cmp_t_LHC-pythia} Comparisons between \POWHEG{}
and \PYTHIA{} results for $t$-channel top production at the LHC $pp$
collider.}
\end{figure}

In figs.~\ref{fig:cmp_t_LHC-mcatnlo} and~\ref{fig:cmp_t_LHC-pythia}
similar results are reported for the LHC $pp$ collider. Only plots for
the $t$-channel production are shown, the $s$-channel process
having a negligible impact at the LHC.

Figure~\ref{fig:cmp_t_LHC-mcatnlo} contains comparisons between
\POWHEG{}, \MCatNLO{} and NLO results. No significant differences with
respect to what we observed at the Tevatron arise in any plot, so that
we refer to the previous section for comments.

In the \PYTHIA{} and \POWHEG{} comparisons shown in
fig.~\ref{fig:cmp_t_LHC-pythia}, we immediately notice that the
\POWHEG{} enhancement of high-$\pt$ tails in panels~(\emph{e}) and~(\emph{f})
is here more marked, even if still small.  This may again 
be related to the lack of matrix-element corrections in
\PYTHIA{}, resulting in larger discrepancies at the LHC with respect to
the Tevatron case.

In panels~(\emph{c}) and~(\emph{e}), one can also notice different low-$\pt$
shapes with respect to the same plots showing the \POWHEG{}+\HERWIG{} results
of fig.~\ref{fig:cmp_t_LHC-mcatnlo}.  We have verified that these differences
are due to the inclusion of multiple interactions~(MI) in the default
\PYTHIA{}.\footnote{These account for events where more than one parton pair
  in the same incoming hadrons give rise to hard interactions.}  If we limit
ourselves to the results without MI (i.e.~setting {\tt MSTP(81)=0} in
\PYTHIA{}), the agreement is much better.

\cleardoublepage

\subsection{Top-quark decay}
As explained in sec.~\ref{sec:top_decay}, in our calculation we have
implemented spin correlations in top decay.  Sizable effects are thus
visible when comparing our results with SMC programs that do not
implement them. \MCatNLO{} accounts for these effects with
approximately the same method that we use. Hence, we expect to have
good agreement with \MCatNLO{} and visible discrepancies when
comparing with \PYTHIA{}.

\begin{figure}[htb] 
\centerline{ 
\subfigure[ $s$-channel]{
\epsfig{figure=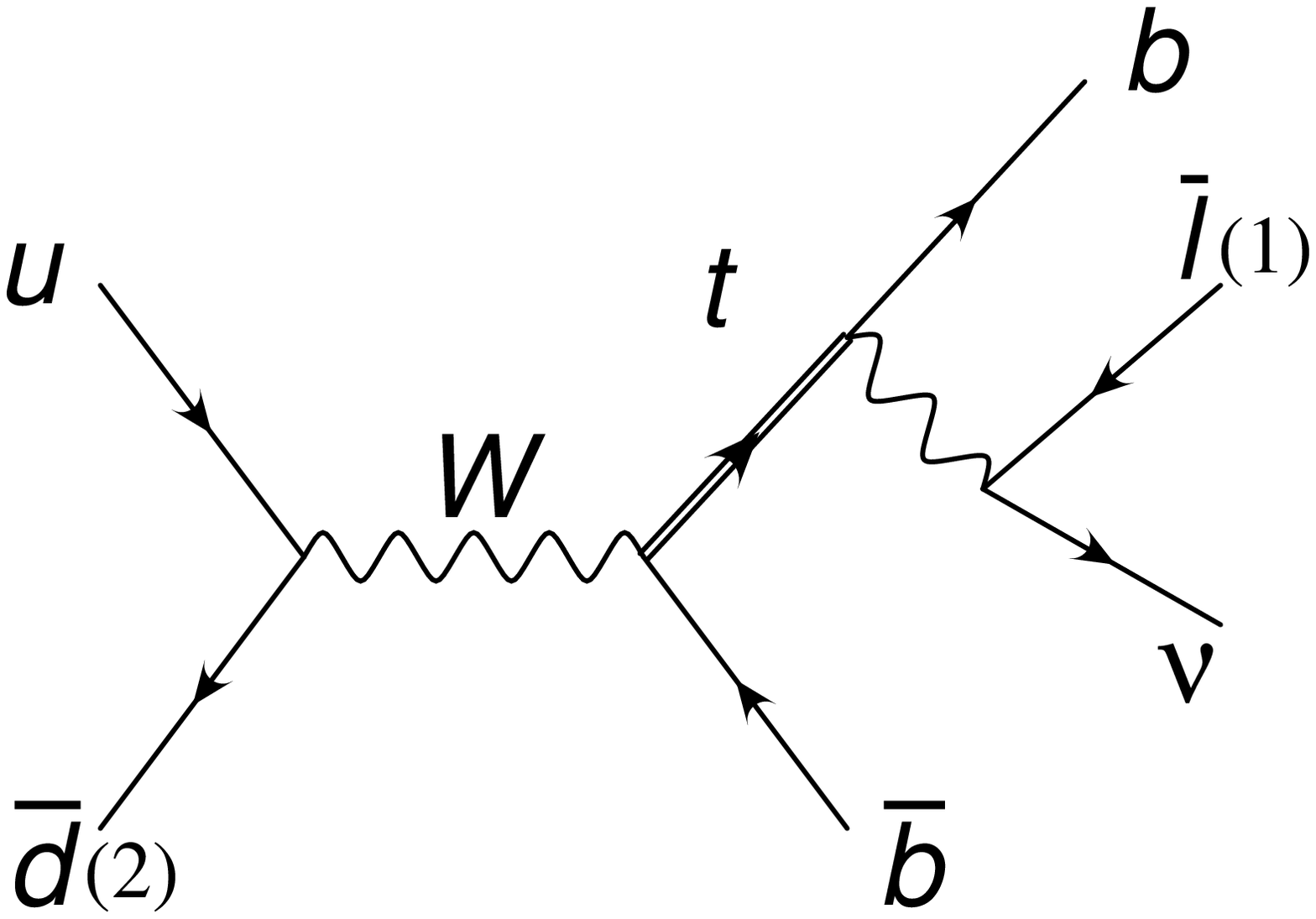,width=0.4\textwidth,clip=}}
\qquad
\subfigure[ $t$-channel]{ 
\epsfig{figure=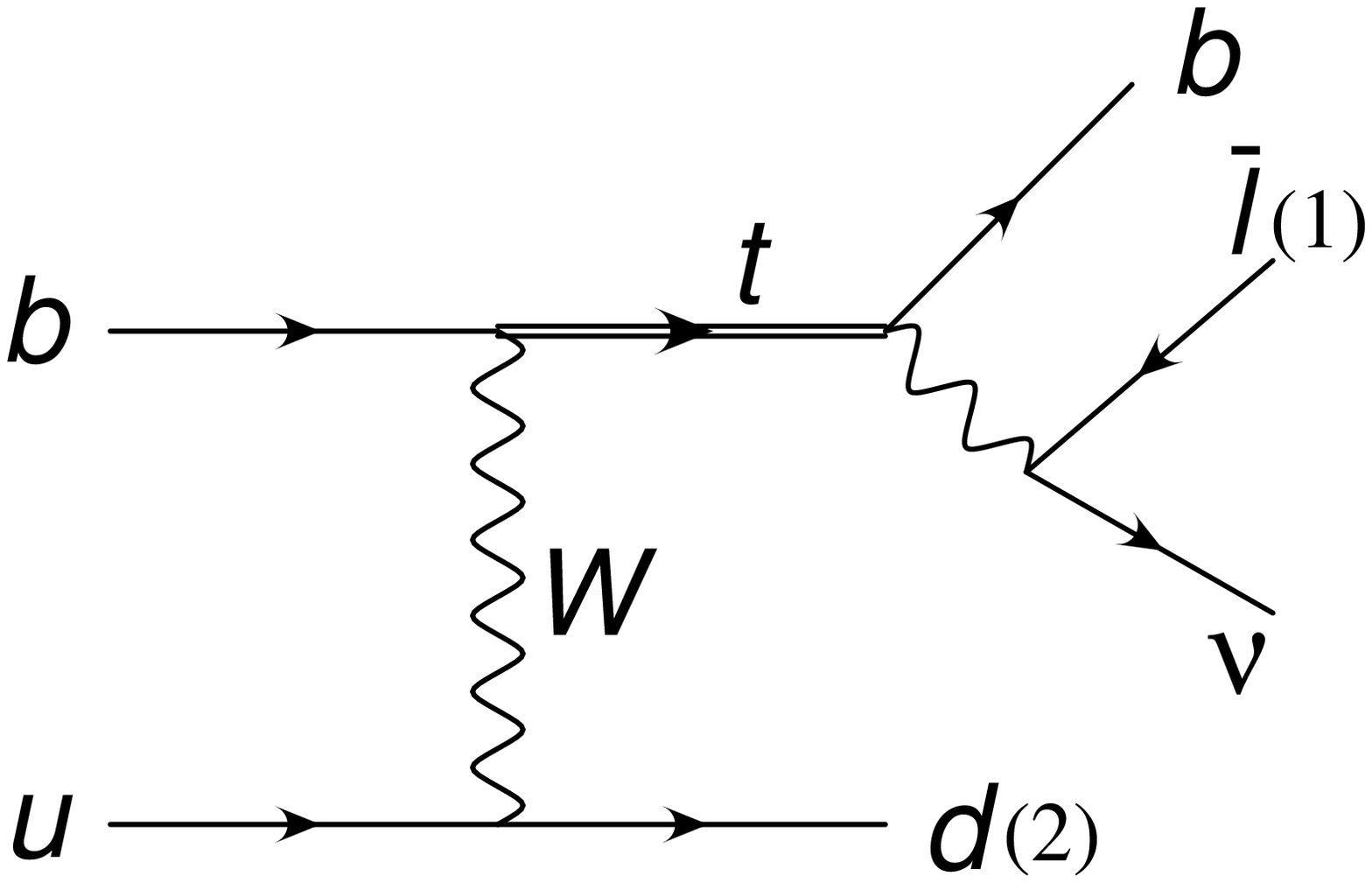,width=0.4\textwidth,clip=}}
} 
\caption{\label{fig:decorr} Lepton $(1)$ and down-type quark
$(2)$ used to study spin correlations in top decay.}
\end{figure}

Due to the V-A structure of the weak current, the best observables to
highlight eventual discrepancies are those involving the angle between the
charged lepton $\bar{\ell}$ coming from top decay and the direction of the
down-type quark entering the $W$ vertex involved in top production, as shown
in fig.~\ref{fig:decorr}.

At the Born level, the down-type quark direction is possibly identified with
the beam axis for $s$-channel production, while, for $t$-channel production,
it often corresponds to the hardest jet axis (see ref.~\cite{Sullivan:2005ar}
for further details).

For sake of comparison, we have set the top virtuality $M^2=m_t^2$ and we
have taken the values $\Gamma_t=1.7$~GeV and $\Gamma_W=2.141$~GeV in the
evaluation of upper bounds of the decay amplitudes in eq.~(\ref{eq:ub_dec})
and in the decayed matrix element $\mathcal{M}^f_{\rm dec}$.  Furthermore, we
have applied cuts similar to those used in ref.~\cite{Frixione:2007zp}, both
for the Tevatron and for the LHC, namely
\begin{eqnarray}
&&\pt^B\ge 20~{\rm GeV}\,,\qquad\quad |\eta^B|\le 2\,,
\\
&&\pt^{\bar{\ell}}\ge 10~{\rm GeV}\,,\qquad\quad |\eta^{\bar{\ell}}|\le 2.5\,,
\\
&&\pt^{\nu}\ge 20~{\rm GeV}\,.
\end{eqnarray}
We denote with the superscript $B$ the top jet, i.e.~the jet that contains
the hardest \mbox{$b$-flavoured} hadron (not the $\bar{b}$).  In single-top
processes, this comes almost exclusively from the bottom quark emerging from
top decay.  In $t$-channel production, in order to isolate a central hardest
light jet, we apply the further cuts
\begin{equation}
\pt^{\,j_1}\ge 20~{\rm GeV}\,,\qquad
|\eta^{\,j_1}|\le 2.5\,.
\end{equation}

\begin{figure}[htb]
\begin{center}
\epsfig{file=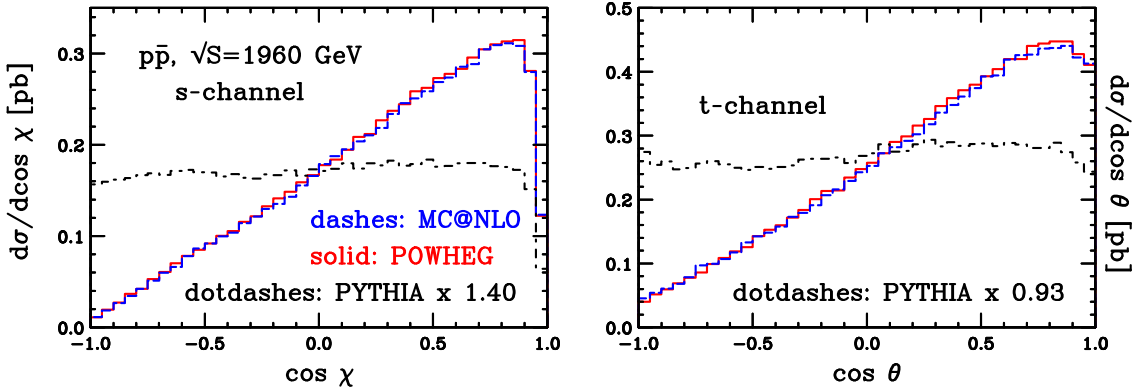,width=\figwidth}
\end{center}
\captskip
\caption{\label{fig:cmp3_st_TeV} Comparisons between \POWHEG{},
\MCatNLO{} and \PYTHIA{} angular correlations for $s$-~(left) and $t$-channel
(right) top production at the Tevatron $p\bar{p}$ collider.}
\end{figure}

In fig.~\ref{fig:cmp3_st_TeV} we show comparisons for the Tevatron
$p\bar{p}$ collider. On the left panel, we plot the $s$-channel
differential cross section as a function of $\cos\chi$, where $\chi$
is the angle between the hardest charged lepton $\bar{\ell}$, which we
assume coming from top decay, and the direction of the incoming
parton with negative rapidity (the $\sminus$ direction of the $z$
axis), as seen in the top rest frame.  Such angle is sensitive to the
spin correlation between $\bar{\ell}$ and the incoming $\bar{d}$
quark, which, at the Tevatron, is pulled out mostly from the
antiproton traveling in the negative direction.  On the right panel,
we plot the $t$-channel differential cross section as a function of
$\cos\theta$, where $\theta$ is the angle between $\bar{\ell}$ and the
hardest jet, always evaluated in the top rest frame. In both plots, we
observe a remarkable good agreement with \MCatNLO{} and the expected
discrepancy with \PYTHIA{}, that only performs a spin-averaged top
decay.

\begin{figure}[htb]
\begin{center}
\epsfig{file=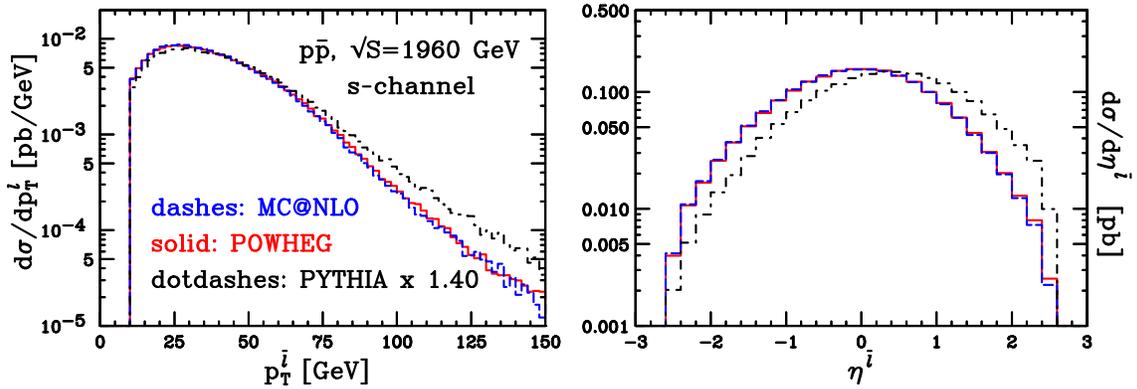,width=\figwidth}
\end{center}
\captskip
\caption{\label{fig:cmp4_s_TeV} Comparisons between \POWHEG{},
\MCatNLO{} and \PYTHIA{} transverse momentum and pseudorapidity of the
lepton coming from the top decay, for $s$-channel top production at
the Tevatron $p\bar{p}$ collider.}
\end{figure}

In fig.~\ref{fig:cmp4_s_TeV} we plot the transverse momentum and
pseudorapidity of the hardest charged lepton, for $s$-channel
production at Tevatron.  The difference between \PYTHIA{} and
\POWHEG{} (or \MCatNLO{}) can be shown to arise because of
spin-correlation effects. To test this, we run \POWHEG{} with an
undecayed top in the final state, leaving \PYTHIA{} to perform the
decay: after rescaling the plots with the appropriate $K$ factor, we
obtain the same behaviour as \PYTHIA{} standalone.

\begin{figure}[htb]
\begin{center}
\epsfig{file=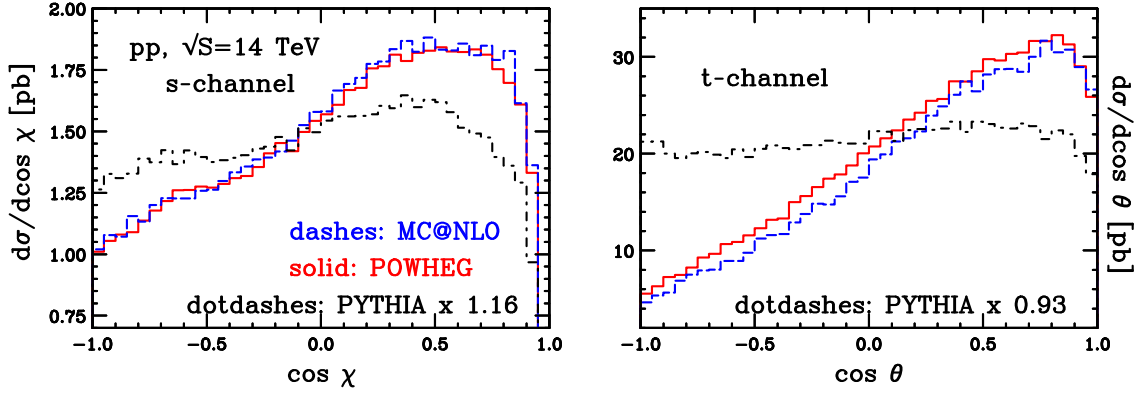,width=\figwidth}
\end{center}
\captskip
\caption{\label{fig:cmp3_st_LHC} Comparisons between \POWHEG{},
\MCatNLO{} and \PYTHIA{} angular correlations for $s$- (left) and $t$-channel
(right) top production at the LHC $pp$ collider.}
\end{figure}

In fig.~\ref{fig:cmp3_st_LHC}, the same distributions of
fig.~\ref{fig:cmp3_st_TeV} are shown for the LHC collider.
The same considerations done for the
Tevatron apply for the LHC results.

\subsection{Dips in the rapidity distributions}
In previous
works~\cite{Nason:2006hfa,Frixione:2007nw,Alioli:2008gx,Alioli:2008tz},
we have extensively discussed the presence of sizable mismatches
between \POWHEG{} and \MCatNLO{} results in the rapidity difference
between the hardest jet and the heavy system recoiling against it.
More specifically, the \MCatNLO{} results exhibit, for this quantity,
a dip at zero rapidity, not visible in \POWHEG{}.  This problem was
originally pointed out in ref.~\cite{Mangano:2006rw} in the framework
of $t\bar{t}$ production, and its origin was traced back to \HERWIG{},
that shows an even deeper dip for the same quantity. In
ref.~\cite{Hamilton:2009za}, in the framework of Higgs production,
this problem and its Shower Monte Carlo origin was accurately studied.

\begin{figure}[htb]
\begin{center}
\epsfig{file=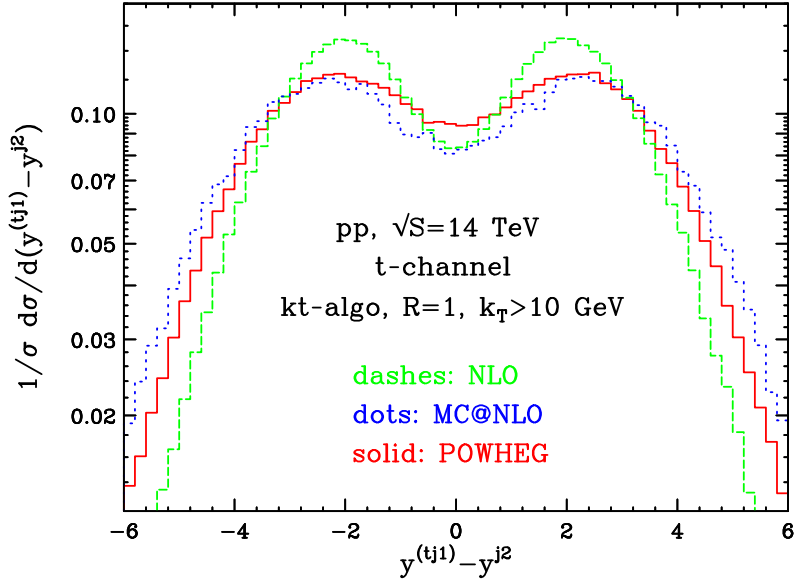,width=0.7\figwidth}
\end{center}
\captskip
\caption{\label{fig:cmpdy_t_LHC} Comparison between \POWHEG{}, \MCatNLO{} and
NLO results for the rapidity difference between the rapidity of the
top-quark--hardest-jet system and the rapidity of the next-to-hardest jet,
for $t$-channel top production at the LHC $pp$ collider. Plots are normalized
to the total cross section.}
\end{figure}

In single-top production, the suitable quantity where to observe this
mismatch is the rapidity difference between the top-quark--hardest-jet system
and the next-to-hardest jet.  As one can see in fig.~\ref{fig:cmpdy_t_LHC},
in this case a dip in the central rapidity region is already present at the
next-to-leading order. This feature may mask an eventual dip in
\MCatNLO{}. In fact, the two showered results are fairly similar, with the
dip being slightly more pronounced in \MCatNLO{}.

In recent talks~\cite{Nason:talk_Firenze,Nason:talk_CERN,Nason:talk_Madison},
one of us proposed a possible explanation of the presence of these dips in
the \MCatNLO{} results. In the following we illustrate this explanation and
show that it is also compatible with the case at hand.

We can schematically represent the \MCatNLO{} cross section for
the hardest emission with the following formula
\begin{eqnarray}
\label{eq:mcnlo_hardemission}
d\sigma &=& \underbrace{\bar{B}^{\rm\sss MC}(\tmmathbf{\bar{\Phi}}_n)\,
  d\bar{\bf \Phi}_n }_{\clS \rm\ event}\
\underbrace{\lq \Delta^{\rm\sss MC}(\tmmathbf{\bar{\Phi}}_n,t_0) + 
\Delta^{\rm\sss MC}(\tmmathbf{\bar{\Phi}}_n,t) \, 
\frac{R^{\rm\sss MC}(\tmmathbf{{\Phi}}_{n+1})}{B(\tmmathbf{\bar{\Phi}}_n)}
\, d\Rad^{\rm\sss MC}  
\rq}_{\rm MC\ shower} \nonumber\\
&& +\ \underbrace{\Big[R(\tmmathbf{{\Phi}}_{n+1}) - R^{\rm\sss
  MC}(\tmmathbf{{\Phi}}_{n+1})\Big]\, 
d\bar{\bf \Phi}_n\, d\Rad^{\rm\sss MC} }_{\clH \rm\ event}\,.
\end{eqnarray}
The terminology ``$\clS$'' and ``$\clH$ events'' is defined
in the original \MCatNLO{} papers~\cite{Frixione:2002ik,Frixione:2003ei}.
We have
\begin{eqnarray}
 \bar{B}^{\rm\sss
  MC}(\tmmathbf{{\bar{\Phi}}}_{n}) &=& B(\tmmathbf{\bar{\Phi}}_n)+ 
V(\tmmathbf{\bar{\Phi}}_n) +\int d\Rad^{\rm\sss MC}\ R^{\rm\sss
  MC}(\tmmathbf{{\Phi}}_{n+1})\,, \label{eq:MCATNLObbarra}\\ 
R^{\rm\sss MC}(\tmmathbf{{\Phi}}_{n+1})&=&B(\tmmathbf{\bar{\Phi}}_n)\
\frac{\as(t)}{2\pi}\frac{1}{t}\ P(z)\,, \\
\Delta^{\rm\sss MC}(\tmmathbf{\bar{\Phi}}_n,t)&=&\exp\lg-\int d\Rad^{\rm\sss MC}\ 
\frac{\as(t)}{2\pi}\frac{1}{t}\ P(z)\,
\theta\(\kt(\tmmathbf{{\Phi}}_{n+1}) - t\)\rg\,,
\end{eqnarray}
where $P(z)$ are the Altarelli-Parisi splitting kernels and
\mbox{$d\Rad=d\Radmc \equiv dz\, dt\, d\phi/(2\pi)$}.  Notice that, on
the right hand side of eq.~(\ref{eq:MCATNLObbarra}), divergent
quantities appear, and only their sum is finite. In the \MCatNLO{}
framework, they are dealt with the subtraction method.

The ``MC shower'' factor in eq.~(\ref{eq:mcnlo_hardemission}) shows that
the hardest emission is produced by running the \HERWIG{} shower Monte Carlo,
starting with the event kinematics $\tmmathbf{{\bar{\Phi}}}_{n}$. In fact,
the Monte Carlo may not generate the hardest radiation as its first emission.
It was shown in ref.~\cite{Nason:2004rx}, however, that
formula~(\ref{eq:mcnlo_hardemission}) does correctly represent the hardest
emission probability up to subleading effects, that we here assume to be
irrelevant for our argument.

In the production of a high-$\pt$ parton, formula~(\ref{eq:mcnlo_hardemission})
yields
\begin{eqnarray}
d\sigma &\approx& \bar{B}^{\rm\sss MC}(\tmmathbf{\bar{\Phi}}_n)\,
\frac{R^{\rm\sss MC}(\tmmathbf{{\Phi}}_{n+1})}{B(\tmmathbf{\bar{\Phi}}_n)}
\ d\bar{\bf \Phi}_n\, d\Rad^{\rm\sss MC} +
\Big[ R(\tmmathbf{{\Phi}}_{n+1}) - R^{\rm\sss MC}(\tmmathbf{{\Phi}}_{n+1})\Big]
\, d\bar{\bf \Phi}_n\, d\Rad^{\rm\sss MC}  \nonumber\\
&\approx& R(\tmmathbf{{\Phi}}_{n+1})\, d\bar{\bf \Phi}_n\, d\Rad^{\rm\sss MC}
+ \underbrace{
\(\frac{\bar{B}^{\rm\sss
    MC}(\tmmathbf{\bar{\Phi}}_n)}{B(\tmmathbf{\bar{\Phi}}_n)}-1\)}_{{{\cal
      O}(\as)}}  R^{\rm\sss MC}(\tmmathbf{{\Phi}}_{n+1})\, d\bar{\bf  
  \Phi}_n\, d\Rad^{\rm\sss MC}\,, 
\end{eqnarray}
where we have used the fact that $\Delta^{\rm\sss
  MC}(\tmmathbf{\bar{\Phi}}_n,t)\approx 1$ in this limit.
The first term correctly describes the hard radiation in the whole phase
space.  The second term, while formally subleading in $\as$, is responsible
for the dip.  In fact, the dip present in \HERWIG{}
propagates here with a weight proportional to $(\bar{B}^{\rm\sss MC}/B-1)$.
Although subleading, this term can be significant for processes with large
$K$ factors.

In the processes studied so far, this ratio was significantly higher than 1
(see, for example, $gg\to H$), so that the effect was particularly
visible. In single-top production,
instead, due to the small NLO $K$ factor, one has $\bar{B}^{\rm\sss
MC}/B\approx 1$.  This, together with the fact that the fixed NLO result
already presents a central dip for $y^{(t\,j_1)}-y^{j_2}$, results in small
discrepancies between \MCatNLO{} and \POWHEG{} (see
fig.~\ref{fig:cmpdy_t_LHC}).

We notice that a similar mechanism (i.e.~via a large $\bar{B}/B$ factor) for
generating large NNLO terms operates also in \POWHEG{}, and has been
discussed in ref.~\cite{Alioli:2008tz} in the framework of Higgs production,
as being responsible for a hard Higgs boson $\pt$ spectrum. In \POWHEG{},
however, this mechanism cannot generate any dip, since here \HERWIG{} has no
role in the generation of the hardest radiation.

\section{Conclusions}
\label{sec:conc}
In this paper we have described a complete implementation of $s$- and
$t$-channel single-top production at next-to-leading order in QCD, in
the \POWHEG{} framework. This is the first \POWHEG{} implementation of
a process where both initial- and final-state radiation is present.
The calculation for top production has been performed within the
Frixione-Kunszt-Signer subtraction
approach~\cite{Frixione:1995ms,Frixione:1997np}, modified according to
ref.~\cite{Frixione:2007vw}.  We accounted for spin-correlation
effects in top-quark decay with a method analogous to the one proposed
in ref.~\cite{Frixione:2007zp}.
The results of our work have been extensively compared with the \MCatNLO{}
and \PYTHIA{} Shower Monte Carlo programs, together with the fixed
next-to-leading order calculation, both for the Tevatron and for the LHC.

The \MCatNLO{} results are in good agreement with \POWHEG{}, also for
quantities sensitive to angular correlations in top decay.

The \PYTHIA{} results, normalized to the total NLO cross section, show
fair agreement with ours for inclusive quantities that do not involve
the top-decay products.
As expected, we have found sizable mismatches with \PYTHIA{} when
considering distributions involving top-decay products, such as
angular-correlation measurements and charged-lepton transverse
momentum and pseudorapidity.
We have also found differences between our results and the \MCatNLO{}
and \PYTHIA{} ones in the hardest $\bar{b}$-flavoured hadron
transverse momentum and rapidity. The high-$\pt$ mismatch with
\PYTHIA{} may be a consequence of the lack of matrix-element
corrections in the latter, while we attribute the low-$\pt$
disagreement with \MCatNLO{} to the sizable difference that we observe
in the rapidity distribution.

The computer code for the \POWHEG{} implementation presented in this
paper is available, together with the manual, at the site
\url{http://moby.mib.infn.it/~nason/POWHEG}.

\section*{Acknowledgments}
We thank Mike Seymour and Torbj\"orn Sj\"ostrand for useful discussions.

\bibliography{paper}

\providecommand{\href}[2]{#2}\begingroup\raggedright\begin{thebibliography}{10}

\bibitem{Aaltonen:2009jj}
{\bf CDF} Collaboration, T.~Aaltonen {\em et~al.}, {\it {First Observation of
  Electroweak Single Top Quark Production}},
  \href{http://xxx.lanl.gov/abs/arXiv:0903.0885}{{\tt arXiv:0903.0885}}.

\bibitem{Abazov:2009ii}
{\bf D0} Collaboration, V.~M. Abazov {\em et~al.}, {\it {Observation of Single
  Top Quark Production}},  \href{http://xxx.lanl.gov/abs/arXiv:0903.0850}{{\tt
  arXiv:0903.0850}}.

\bibitem{Beneke:2000hk}
M.~Beneke {\em et~al.}, {\it {Top quark physics}},
  \href{http://xxx.lanl.gov/abs/hep-ph/0003033}{{\tt hep-ph/0003033}}.

\bibitem{Harris:2002md}
B.~W. Harris, E.~Laenen, L.~Phaf, Z.~Sullivan, and S.~Weinzierl, {\it {The
  Fully differential single top quark cross-section in next to leading order
  QCD}},  {\em Phys. Rev.} {\bf D66} (2002) 054024,
  [\href{http://xxx.lanl.gov/abs/hep-ph/0207055}{{\tt hep-ph/0207055}}].

\bibitem{Alwall:2006bx}
J.~Alwall {\em et~al.}, {\it {Is $V_{tb} = 1$?}},  {\em Eur. Phys. J.} {\bf
  C49} (2007) 791--801, [\href{http://xxx.lanl.gov/abs/hep-ph/0607115}{{\tt
  hep-ph/0607115}}].

\bibitem{Mahlon:1996pn}
G.~Mahlon and S.~J. Parke, {\it {Improved spin basis for angular correlation
  studies in single top quark production at the Tevatron}},  {\em Phys. Rev.}
  {\bf D55} (1997) 7249--7254,
  [\href{http://xxx.lanl.gov/abs/hep-ph/9611367}{{\tt hep-ph/9611367}}].

\bibitem{Mahlon:1999gz}
G.~Mahlon and S.~J. Parke, {\it {Single top quark production at the LHC:
  Understanding spin}},  {\em Phys. Lett.} {\bf B476} (2000) 323--330,
  [\href{http://xxx.lanl.gov/abs/hep-ph/9912458}{{\tt hep-ph/9912458}}].

\bibitem{Tait:2000sh}
T.~M.~P. Tait and C.~P. Yuan, {\it {Single top quark production as a window to
  physics beyond the standard model}},  {\em Phys. Rev.} {\bf D63} (2001)
  014018, [\href{http://xxx.lanl.gov/abs/hep-ph/0007298}{{\tt
  hep-ph/0007298}}].

\bibitem{Cao:2007ea}
Q.-H. Cao, J.~Wudka, and C.~P. Yuan, {\it {Search for New Physics via Single
  Top Production at the LHC}},  {\em Phys. Lett.} {\bf B658} (2007) 50--56,
  [\href{http://xxx.lanl.gov/abs/arXiv:0704.2809}{{\tt arXiv:0704.2809}}].

\bibitem{Plehn:2009it}
T.~Plehn, M.~Rauch, and M.~Spannowsky, {\it {Understanding Single Tops using
  Jets}},  \href{http://xxx.lanl.gov/abs/arXiv:0906.1803}{{\tt
  arXiv:0906.1803}}.

\bibitem{Bordes:1994ki}
G.~Bordes and B.~van Eijk, {\it {Calculating QCD corrections to single top
  production in hadronic interactions}},  {\em Nucl. Phys.} {\bf B435} (1995)
  23--58.

\bibitem{Stelzer:1997ns}
T.~Stelzer, Z.~Sullivan, and S.~Willenbrock, {\it {Single top quark production
  via $W$-gluon fusion at next-to-leading order}},  {\em Phys. Rev.} {\bf D56}
  (1997) 5919--5927, [\href{http://xxx.lanl.gov/abs/hep-ph/9705398}{{\tt
  hep-ph/9705398}}].

\bibitem{Sullivan:2004ie}
Z.~Sullivan, {\it {Understanding single-top-quark production and jets at hadron
  colliders}},  {\em Phys. Rev.} {\bf D70} (2004) 114012,
  [\href{http://xxx.lanl.gov/abs/hep-ph/0408049}{{\tt hep-ph/0408049}}].

\bibitem{Campbell:2004ch}
J.~M. Campbell, R.~K. Ellis, and F.~Tramontano, {\it {Single top production and
  decay at next-to-leading order}},  {\em Phys. Rev.} {\bf D70} (2004) 094012,
  [\href{http://xxx.lanl.gov/abs/hep-ph/0408158}{{\tt hep-ph/0408158}}].

\bibitem{Campbell:2005bb}
J.~M. Campbell and F.~Tramontano, {\it {Next-to-leading order corrections to
  $Wt$ production and decay}},  {\em Nucl. Phys.} {\bf B726} (2005) 109--130,
  [\href{http://xxx.lanl.gov/abs/hep-ph/0506289}{{\tt hep-ph/0506289}}].

\bibitem{Cao:2004ap}
Q.-H. Cao, R.~Schwienhorst, and C.~P. Yuan, {\it {Next-to-leading order
  corrections to single top quark production and decay at Tevatron. 1.
  $s$-channel process}},  {\em Phys. Rev.} {\bf D71} (2005) 054023,
  [\href{http://xxx.lanl.gov/abs/hep-ph/0409040}{{\tt hep-ph/0409040}}].

\bibitem{Cao:2005pq}
Q.-H. Cao, R.~Schwienhorst, J.~A. Benitez, R.~Brock, and C.~P. Yuan, {\it
  {Next-to-leading order corrections to single top quark production and decay
  at the Tevatron: 2. $t$-channel process}},  {\em Phys. Rev.} {\bf D72} (2005)
  094027, [\href{http://xxx.lanl.gov/abs/hep-ph/0504230}{{\tt
  hep-ph/0504230}}].

\bibitem{Campbell:2009ss}
J.~M. Campbell, R.~Frederix, F.~Maltoni, and F.~Tramontano, {\it {$t$-channel
  single-top production at hadron colliders}},
  \href{http://xxx.lanl.gov/abs/arXiv:0903.0005}{{\tt arXiv:0903.0005}}.

\bibitem{Frixione:2002ik}
S.~Frixione and B.~R. Webber, {\it {Matching NLO QCD computations and parton
  shower simulations}},  {\em JHEP} {\bf 06} (2002) 029,
  [\href{http://xxx.lanl.gov/abs/hep-ph/0204244}{{\tt hep-ph/0204244}}].

\bibitem{Frixione:2003ei}
S.~Frixione, P.~Nason, and B.~R. Webber, {\it {Matching NLO QCD and parton
  showers in heavy flavour production}},  {\em JHEP} {\bf 08} (2003) 007,
  [\href{http://xxx.lanl.gov/abs/hep-ph/0305252}{{\tt hep-ph/0305252}}].

\bibitem{Frixione:2005vw}
S.~Frixione, E.~Laenen, P.~Motylinski, and B.~R. Webber, {\it {Single-top
  production in MC@NLO}},  {\em JHEP} {\bf 03} (2006) 092,
  [\href{http://xxx.lanl.gov/abs/hep-ph/0512250}{{\tt hep-ph/0512250}}].

\bibitem{Frixione:2008yi}
S.~Frixione, E.~Laenen, P.~Motylinski, B.~R. Webber, and C.~D. White, {\it
  {Single-top hadroproduction in association with a W boson}},  {\em JHEP} {\bf
  07} (2008) 029, [\href{http://xxx.lanl.gov/abs/arXiv:0805.3067}{{\tt
  arXiv:0805.3067}}].

\bibitem{Nason:2004rx}
P.~Nason, {\it {A new method for combining NLO QCD with shower Monte Carlo
  algorithms}},  {\em JHEP} {\bf 11} (2004) 040,
  [\href{http://xxx.lanl.gov/abs/hep-ph/0409146}{{\tt hep-ph/0409146}}].

\bibitem{Frixione:2007vw}
S.~Frixione, P.~Nason, and C.~Oleari, {\it {Matching NLO QCD computations with
  Parton Shower simulations: the POWHEG method}},  {\em JHEP} {\bf 11} (2007)
  070, [\href{http://xxx.lanl.gov/abs/arXiv:0709.2092}{{\tt arXiv:0709.2092}}].

\bibitem{Nason:2006hfa}
P.~Nason and G.~Ridolfi, {\it {A positive-weight next-to-leading-order Monte
  Carlo for $Z$ pair hadroproduction}},  {\em JHEP} {\bf 08} (2006) 077,
  [\href{http://xxx.lanl.gov/abs/hep-ph/0606275}{{\tt hep-ph/0606275}}].

\bibitem{Frixione:2007nw}
S.~Frixione, P.~Nason, and G.~Ridolfi, {\it {A Positive-Weight
  Next-to-Leading-Order Monte Carlo for Heavy Flavour Hadroproduction}},  {\em
  JHEP} {\bf 09} (2007) 126,
  [\href{http://xxx.lanl.gov/abs/arXiv:0707.3088}{{\tt arXiv:0707.3088}}].

\bibitem{LatundeDada:2006gx}
O.~Latunde-Dada, S.~Gieseke, and B.~Webber, {\it {A positive-weight
  next-to-leading-order Monte Carlo for $e^+ e^-$ annihilation to hadrons}},
  {\em JHEP} {\bf 02} (2007) 051,
  [\href{http://xxx.lanl.gov/abs/hep-ph/0612281}{{\tt hep-ph/0612281}}].

\bibitem{LatundeDada:2008bv}
O.~Latunde-Dada, {\it {Applying the POWHEG method to top pair production and
  decays at the ILC}},  {\em Eur. Phys. J.} {\bf C58} (2008) 543--554,
  [\href{http://xxx.lanl.gov/abs/arXiv:0806.4560}{{\tt arXiv:0806.4560}}].

\bibitem{Alioli:2008gx}
S.~Alioli, P.~Nason, C.~Oleari, and E.~Re, {\it {NLO vector-boson production
  matched with shower in POWHEG}},  {\em JHEP} {\bf 07} (2008) 060,
  [\href{http://xxx.lanl.gov/abs/arXiv:0805.4802}{{\tt arXiv:0805.4802}}].

\bibitem{Hamilton:2008pd}
K.~Hamilton, P.~Richardson, and J.~Tully, {\it {A Positive-Weight
  Next-to-Leading Order Monte Carlo Simulation of Drell-Yan Vector Boson
  Production}},  {\em JHEP} {\bf 10} (2008) 015,
  [\href{http://xxx.lanl.gov/abs/arXiv:0806.0290}{{\tt arXiv:0806.0290}}].

\bibitem{Papaefstathiou:2009sr}
A.~Papaefstathiou and O.~Latunde-Dada, {\it {NLO production of $W'$ bosons at
  hadron colliders using the MC@NLO and POWHEG methods}},
  \href{http://xxx.lanl.gov/abs/arXiv:0901.3685}{{\tt arXiv:0901.3685}}.

\bibitem{Alioli:2008tz}
S.~Alioli, P.~Nason, C.~Oleari, and E.~Re, {\it {NLO Higgs boson production via
  gluon fusion matched with shower in POWHEG}},  {\em JHEP} {\bf 04} (2009)
  002, [\href{http://xxx.lanl.gov/abs/arXiv:0812.0578}{{\tt arXiv:0812.0578}}].

\bibitem{Hamilton:2009za}
K.~Hamilton, P.~Richardson, and J.~Tully, {\it {A Positive-Weight
  Next-to-Leading Order Monte Carlo Simulation for Higgs Boson Production}},
  {\em JHEP} {\bf 04} (2009) 116,
  [\href{http://xxx.lanl.gov/abs/arXiv:0903.4345}{{\tt arXiv:0903.4345}}].

\bibitem{Corcella:2000bw}
G.~Corcella {\em et~al.}, {\it {HERWIG 6: An event generator for hadron
  emission reactions with interfering gluons (including supersymmetric
  processes)}},  {\em JHEP} {\bf 01} (2001) 010,
  [\href{http://xxx.lanl.gov/abs/hep-ph/0011363}{{\tt hep-ph/0011363}}].

\bibitem{Corcella:2002jc}
G.~Corcella {\em et~al.}, {\it Herwig 6.5 release note},
  \href{http://xxx.lanl.gov/abs/hep-ph/0210213}{{\tt hep-ph/0210213}}.

\bibitem{Sjostrand:2006za}
T.~Sjostrand, S.~Mrenna, and P.~Skands, {\it Pythia 6.4 physics and manual},
  {\em JHEP} {\bf 05} (2006) 026,
  [\href{http://xxx.lanl.gov/abs/hep-ph/0603175}{{\tt hep-ph/0603175}}].

\bibitem{Frixione:2007zp}
S.~Frixione, E.~Laenen, P.~Motylinski, and B.~R. Webber, {\it {Angular
  correlations of lepton pairs from vector boson and top quark decays in Monte
  Carlo simulations}},  {\em JHEP} {\bf 04} (2007) 081,
  [\href{http://xxx.lanl.gov/abs/hep-ph/0702198}{{\tt hep-ph/0702198}}].

\bibitem{Frixione:1995ms}
S.~Frixione, Z.~Kunszt, and A.~Signer, {\it {Three-jet cross sections to
  next-to-leading order}},  {\em Nucl. Phys.} {\bf B467} (1996) 399--442,
  [\href{http://xxx.lanl.gov/abs/hep-ph/9512328}{{\tt hep-ph/9512328}}].

\bibitem{Frixione:1997np}
S.~Frixione, {\it {A general approach to jet cross sections in QCD}},  {\em
  Nucl. Phys.} {\bf B507} (1997) 295--314,
  [\href{http://xxx.lanl.gov/abs/hep-ph/9706545}{{\tt hep-ph/9706545}}].

\bibitem{Alwall:2007st}
J.~Alwall {\em et~al.}, {\it {MadGraph/MadEvent v4: The New Web Generation}},
  {\em JHEP} {\bf 09} (2007) 028,
  [\href{http://xxx.lanl.gov/abs/arXiv:0706.2334}{{\tt arXiv:0706.2334}}].

\bibitem{MCFM}
\texttt{http://mcfm.fnal.gov}.

\bibitem{ZTOP}
\texttt{http://home.fnal.gov/$\sim$zack/ZTOP/ZTOP.html}.

\bibitem{Nason:2007vt}
P.~Nason, {\it {MINT: a Computer Program for Adaptive Monte Carlo Integration
  and Generation of Unweighted Distributions}},
  \href{http://xxx.lanl.gov/abs/arXiv:0709.2085}{{\tt arXiv:0709.2085}}.

\bibitem{Pumplin:2002vw}
J.~Pumplin {\em et~al.}, {\it {New generation of parton distributions with
  uncertainties from global QCD analysis}},  {\em JHEP} {\bf 07} (2002) 012,
  [\href{http://xxx.lanl.gov/abs/hep-ph/0201195}{{\tt hep-ph/0201195}}].

\bibitem{Catani:1993hr}
S.~Catani, Y.~L. Dokshitzer, M.~H. Seymour, and B.~R. Webber, {\it
  {Longitudinally invariant $k_T$ clustering algorithms for hadron-hadron
  collisions}},  {\em Nucl. Phys.} {\bf B406} (1993) 187--224.

\bibitem{Cacciari:2005hq}
M.~Cacciari and G.~P. Salam, {\it {Dispelling the $N^3$ myth for the $k_T$
  jet-finder}},  {\em Phys. Lett.} {\bf B641} (2006) 57--61,
  [\href{http://xxx.lanl.gov/abs/hep-ph/0512210}{{\tt hep-ph/0512210}}].

\bibitem{Sullivan:2005ar}
Z.~Sullivan, {\it {Angular correlations in single-top-quark and Wjj production
  at next-to-leading order}},  {\em Phys. Rev.} {\bf D72} (2005) 094034,
  [\href{http://xxx.lanl.gov/abs/hep-ph/0510224}{{\tt hep-ph/0510224}}].

\bibitem{Mangano:2006rw}
M.~L. Mangano, M.~Moretti, F.~Piccinini, and M.~Treccani, {\it Matching matrix
  elements and shower evolution for top-quark production in hadronic
  collisions},  {\em JHEP} {\bf 01} (2007) 013,
  [\href{http://xxx.lanl.gov/abs/hep-ph/0611129}{{\tt hep-ph/0611129}}].

\bibitem{Nason:talk_Firenze}
P.~Nason, {\it {Shower Monte Carlo at Next-to-Leading Order}},
  \href{http://xxx.lanl.gov/abs/{http://theory.fi.infn.it/research/nason.pdf}}%
{{\tt {http://theory.fi.infn.it/research/nason.pdf}}}. ~\\Talk given at the
  Universit\`a degli Studi di Firenze, Florence, Italy, 2009.

\bibitem{Nason:talk_CERN}
P.~Nason, {\it {MC at NLO tools}},
  \href{http://xxx.lanl.gov/abs/{http://indico.cern.ch/getFile.py/access?contr%
ibId=2\&resId=0\&materialId=\\slides\&confId=49675}}{{\tt
  {http://indico.cern.ch/getFile.py/access?contribId=2\&resId=0\&materialId=\\%
slides\&confId=49675}}}. ~\\Talk given at {\sl MC4LHC Meeting}, CERN,
  Switzerland, 2009.

\bibitem{Nason:talk_Madison}
P.~Nason, {\it {POWHEG}},
  \href{http://xxx.lanl.gov/abs/{http://agenda.hep.wisc.edu/materialDisplay.py%
?contribId=13\&materialId=\\slides\&confId=189}}{{\tt
  {http://agenda.hep.wisc.edu/materialDisplay.py?contribId=13\&materialId=\\sl%
ides\&confId=189}}}. ~\\Talk given at {\sl LoopFest Symposium}, Madison, WI,
  USA, 2009.

\end{thebibliography}\endgroup

\end{document}